\documentclass[12pt]{article}

\usepackage{amsfonts}
\usepackage{amssymb}
\usepackage{amsmath}
\usepackage{eqsection}
\usepackage{cite}

\newcommand{\beq}{\begin{equation}}
\newcommand{\eeq}{\end{equation}}

\DeclareMathOperator{\sn}{sn}
\DeclareMathOperator{\cn}{cn}
\DeclareMathOperator{\dn}{dn}
\DeclareMathOperator{\jacobisc}{sc}
\DeclareMathOperator{\nd}{nd}

\newcommand{\cu}[1]{{\chi}^{({#1})}}
\newcommand{\ch}[1]{{\hat\chi}^{({#1})}}
\newcommand{\chiso}[1]{{\hat\chi}^{({#1})}_{\text{iso}}}
\newcommand{\Ch}[1]{{\hat C}^{({#1})}}

\def\half{{\tfrac{1}{2}}}

\def\aam#1#2#3 {{\em Adv. Appl. Math.} {\bf#1} (#2) #3}
\def\ap#1#2#3 {{\em Ann. Phys. (NY)} {\bf#1} (#2) #3}
\def\apj#1#2#3 {{\em Astrophys. J.} {\bf#1} (#2) #3}
\def\apjl#1#2#3 {{\em Astrophys. J. Lett.} {\bf#1} (#2) #3}
\def\app#1#2#3 {{\em Acta. Phys. Pol.} {\bf#1} (#2) #3}
\def\ar#1#2#3 {{\em Ann. Rev. Nucl. Part. Sci.} {\bf#1} (#2) #3}
\def\cmp#1#2#3 {{\em Comm. Math. Phys.} {\bf#1} (#2) #3}
\def\cpc#1#2#3 {{\em Computer Phys. Comm.} {\bf#1} (#2) #3}
\def\err#1#2#3 {{\it Erratum} {\bf#1} (#2) #3}
\def\ib#1#2#3 {{\it ibid.} {\bf#1} (#2) #3}
\def\jmp#1#2#3 {{\em J. Math. Phys.} {\bf#1} (#2) #3}
\def\ijmp#1#2#3 {{\em Int. J. Mod. Phys.} {\bf#1} (#2) #3}
\def\jalg#1#2#3 {{\em J. Algebra} {\bf#1} (#2) #3}
\def\jetp#1#2#3 {{\em JETP Lett.} {\bf#1} (#2) #3}
\def\jpa#1#2#3 {{\em J. Phys. A: Math. Gen.} {\bf#1} (#2) #3}
\def\jpg#1#2#3 {{\em J. Phys. G.} {\bf#1} (#2) #3}
\def\jsp#1#2#3 {{\em J. Stat. Phys.} {\bf#1} (#2) #3}
\def\mpl#1#2#3 {{\em Mod. Phys. Lett.} {\bf#1} (#2) #3}
\def\nat#1#2#3 {{\em Nature (London)} {\bf#1} (#2) #3}
\def\nc#1#2#3 {{\em Nuovo Cim.} {\bf#1} (#2) #3}
\def\nim#1#2#3 {{\em Nucl. Instr. Meth.} {\bf#1} (#2) #3}
\def\np#1#2#3 {{\em Nucl. Phys.} {\bf#1} (#2) #3}
\def\pcps#1#2#3 {{\em Proc. Cam. Phil. Soc.} {\bf#1} (#2) #3}
\def\pja#1#2#3 {{\em Proc. Japan Acad.} {\bf#1} (#2) #3}
\def\pl#1#2#3 {{\em Phys. Lett.} {\bf#1} (#2) #3}
\def\prep#1#2#3 {{\em Phys. Rep.} {\bf#1} (#2) #3}
\def\prev#1#2#3 {{\em Phys. Rev.} {\bf#1} (#2) #3}
\def\preb#1#2#3 {{\em Phys. Rev. B} {\bf#1} (#2) #3}
\def\pree#1#2#3 {{\em Phys. Rev. E} {\bf#1} (#2) #3}
\def\prl#1#2#3 {{\em Phys. Rev. Lett.} {\bf#1} (#2) #3}
\def\prs#1#2#3 {{\em Proc. Roy. Soc.} {\bf#1} (#2) #3}
\def\ptp#1#2#3 {{\em Prog. Theor. Phys.} {\bf#1} (#2) #3}
\def\psa#1#2#3 {{\em Physica Scripta} {\bf#1} (#2) #3}
\def\ps#1#2#3 {{\em Physica} {\bf#1} (#2) #3}
\def\rmp#1#2#3 {{\em Rev. Mod. Phys.} {\bf#1} (#2) #3}
\def\rpp#1#2#3 {{\em Rep. Prog. Phys.} {\bf#1} (#2) #3}
\def\sjnp#1#2#3 {{\em Sov. J. Nucl. Phys.} {\bf#1} (#2) #3}
\def\spj#1#2#3 {{\em Sov. Phys. JETP} {\bf#1} (#2) #3}
\def\spu#1#2#3 {{\em Sov. Phys. Usp.} {\bf#1} (#2) #3}
\def\zp#1#2#3 {{\em Zeit. Phys.} {\bf#1} (#2) #3}

\begin{document}
\title{The susceptibility of the square lattice Ising model:\\
New developments}
\author{W. P. Orrick$^\dag$, B. Nickel$^\S$,
 A. J. Guttmann$^\dag$ \\
and \\
 J. H. H. Perk$^\P$
\vspace{0.2in}\\
$^\dag$Department of Mathematics \& Statistics,\\
The University of Melbourne\\
Parkville, Vic. 3052, Australia
\vspace{0.1in}\\
$^\S$Department of Physics,\\
University of Guelph,\\
Guelph, Ontario, Canada N1G 2W1
\vspace{0.1in}\\
$^\P$Department of Physics,\\ Oklahoma State University, \\
Stillwater, Oklahoma 74078-3072, U.S.A.}
\date{\today}
\maketitle
\bibliographystyle{plain}
\begin{abstract}
We have made substantial advances in elucidating the
properties of the susceptibility of the square lattice
Ising model. We discuss its analyticity properties,
certain closed form expressions for subsets of the coefficients,
and give an algorithm of complexity {\rm O}$(N^6)$ to
determine its first $N$ coefficients. As a result, we have generated
and analyzed series with more than 300 terms in both the high- and 
low-temperature regime. 
We quantify the effect of irrelevant variables
to the scaling-amplitude functions.
In particular, we find and quantify the breakdown of simple scaling,
in the absence of irrelevant scaling fields, arising first at order
$|T-T_c|^{9/4}$, though high-low temperature symmetry is still
preserved. At terms of order $|T-T_c|^{17/4}$ and beyond, this
symmetry is no longer present. The short-distance terms are shown to
have the form $(T-T_c)^p(\log|T-T_c|)^q$ with $p\ge q^2$.
Conjectured exact expressions
for some correlation functions and series coefficients in terms of
elliptic theta functions also foreshadow future developments.
\end{abstract}

\section*{Keywords}
Ising susceptibility, high-temperature series, low-temperature series, scaling
function, irrelevant variables, differentiably finite functions, scaling
fields.
\section{Introduction}
Since Onsager's \cite{onsa} celebrated solution of the Ising model free energy
in 1944, followed by Yang's \cite{yang} proof of Onsager's result for the
spontaneous magnetization in 1952, almost half a century has passed during
which time many, if not most of the world's most able
mathematical physicists have devoted themselves to the problem of elucidating
the susceptibility. Attempting to list all these contributions would produce a
bibliography of prohibitive length, and one that would inevitably commit many
sins of omission. Therefore rather than attempt this, we will only make
mention of those papers that have directly motivated our work here, and
crave the forgiveness of those who we have inadvertently offended.

While much of the notation for describing the square lattice
Ising model is standard, we begin by defining our notation here both
for the benefit of the more casual reader and to emphasize those
cases where we deviate from convention. The interactions in the two
perpendicular directions are taken to be 
\begin{equation}
K=\beta J,\qquad K'=\beta J'.
\end{equation}
but we also often set $K'=K$ to discuss the isotropic lattice. 
For high temperatures, $s=\sinh 2K$ and $s'=\sinh 2K'$ are
appropriate variables for series
expansions \cite{nica}, while for low temperatures, we use $1/s$ and 
$1/s'$ instead.
Thus,
\begin{equation}
s^*=\sinh 2K^* = 1/\sinh 2K = 1/s,\qquad {s'}^*=\sinh 2{K'}^*=1/\sinh 2K'=1/s'.
\end{equation}
In many cases, high-temperature and low-temperature formulas can be 
obtained
from each other by Kramers-Wannier duality with a simple interchange of
primes and stars.
The critical temperature is defined by the condition $s'=s^*$.

A conventional high-temperature variable is $v=\tanh{K}$,
while an often-used low-temper\-a\-ture variable is $u=\exp(-4K)$.
The translations between these and our variables are
\begin{equation}
s=\sinh{2K}=2v/(1-v^2)
\end{equation}
and
\begin{equation}
s^*=2u^{1/2}/(1-u),
\end{equation}
and similarly for the primed variables.

In studying the critical behavior, we will use both the variable
$t = 1 - T_c/T$ and more frequently 
\begin{equation}
\tau = (1/s - s)/2
\text{ (isotropic)}
\end{equation}
to parameterize deviations from the critical temperature. To leading
order, $\tau = 2K_c\sqrt{2}t.$

An elliptic parameterization will be useful, and to that end we define
the elliptic modulus,
\begin{equation}\label{mod}
k=\begin{cases} s'/s^*=ss' &\text{for $T>T_c$,} \\
s^*/s'=1/ss' &\text{for $T<T_c$.}\end{cases}
\end{equation}
Let $\sigma_{i,j}$ be the spin at lattice site $(i,j)$ and
define the two-point function
\begin{equation}
C(M,N) = \langle \sigma_{0,0}\sigma_{M,N}\rangle.
\end{equation}
In terms of the correlation functions the susceptibility is
\begin{equation}\label{1.8}
 \beta^{-1}\chi = \sum \sum (C(M,N)-{\mathcal M}^2)
\end{equation}
where $\mathcal M$ is the magnetization.

In 1956 Syozi and Naya~\cite{SN56} presented an approximation to the
anisotropic high-tempera\@-ture susceptibility
which gave the correct critical point, correct critical exponent,
an amplitude estimate that
was wrong by less than 1\%, and reproduced the first 8 terms of the series
expansion. It was also exact along the disorder line\footnote{For the fully
anisotropic triangular lattice Ising model, with coupling constants
$v_i=\tanh{J_i/kT},$ $i=1,2,3,$ the condition for the disorder line is that
$v_1 v_2 + v_3 = 0.$ Along this line the correlations decay exponentially,
and the partition function factorizes \cite{step70}.
The disorder condition is of no particular
interest in the case of the nearest neighbor square lattice Ising model.}
In 1976 a celebrated paper by Wu, McCoy, Tracy and Barouch \cite{wmtb}
showed how the high- and low-temperature expansions of the susceptibility
could be understood in terms of a multi-particle expansion, with an odd number
of particles being appropriate at high temperatures and an even number at
low temperatures. With this interpretation it became clear that the result
of \cite{SN56} was just the lowest order, or one-particle approximation
to the susceptibility.

In terms of the elliptic modulus $k$~(\ref{mod})
the high- and low-temperature susceptibilities can be written
\begin{equation}\label{1a}
\beta^{-1}\chi_+ =k^{-\frac{1}{2}}(1-k^2)^{\frac{1}{4}} 
\sum_{l=0}^\infty\hat{\chi}^{(2l+1)}
\end{equation}
and
\begin{equation}\label{1b}
\beta^{-1}\chi_- =(1-k^2)^{\frac{1}{4}}
 \sum_{l=1}^\infty\hat{\chi}^{(2l)}
\end{equation}
respectively, where
$\hat{\chi}^{(j)} $ is the sum over all lattice separations of the $j$-particle
contribution to the two-point function, and was first given
\cite{wmtb} as a $2j$-fold multiple integral. It was subsequently shown
that this integral can be reduced to a $j$-fold integral of the form
\begin{equation}\label{e1}
\hat{\chi}^{(j)}=\frac{k^{j/2}}{(2\pi)^j j!}\int du_1 \cdots \int du_j
 (G^{(j)})^2 f^{(j)},
\end{equation}
where $G^{(j)}$ is a fermionic determinant and $f^{(j)}$ is an algebraic
function.
This reduction has been achieved by various
routes~\cite{napp,pt,yamb,yamd,nica,nicb}.
The factor, $G^{(j)}$, appearing in the
integrand, which can be expressed in terms of Pfaffians~\cite{napp},
has been found to have a product form, first by Palmer and Tracy in the
low temperature regime~\cite{pt}, and then independently by Yamada in
both the high- and low-temperature 
regimes~\cite{yamb,yamd}. 
From the product form it readily follows that the first non-zero
term in (\ref{e1}) is
$2^{j(1-j)}k^{j^2/2}.$ From the original expression
one could only conclude that each integral entered at order $k^{j/2},$ 
so clearly massive cancellations occur. While this comes as a surprise if
handling the integral~\cite{wmtb} directly, it follows
straightforwardly~\cite{nicb} from the product form.

In terms of the elliptic modulus, the first terms in the high- and
low-temperature expansions for the isotropic ($K=K'$) susceptibility are
\begin{equation}\label{n1}
\chiso{1} = \frac{k^{\frac{1}{2}}}{(1-k^{\frac{1}{2}})^2} 
\end{equation}
and
\begin{equation}\label{n2}
\chiso{2} = \frac{(1+k^2){\rm E} - (1-k^2){\rm K}}{3\pi(1-k)(1-k^2)},
\end{equation}
where E and K are the complete elliptic integrals\footnote{We 
trust there is no confusion with the coupling constant $K$.}
of the second and first
kind respectively. Anisotropic versions of these formulae are given
in section~\ref{sect:integrals}. Simple forms for higher terms in the
expansion are not known.

In \cite{GE96} one of us gave compelling evidence
that unlike the free-energy and spontaneous
magnetization, the {\em anisotropic} susceptibility $\chi(K,K')$ is not
{\em differentiably finite.} 
A series in $n$ variables, $f({\bf z})$, is said to be
{\em differentiably finite} or {\em D-finite}
if and only if it satisfies a system of $n$ partial differential
equations of the form
\begin{equation}
P_{i,0}({\bf z})f({\bf z}) + P_{i,1}({\bf z})\frac{\partial}{\partial z_i}
f({\bf z}) +\cdots +
P_{i,k_i}({\bf z})\frac{\partial^{k_i}}{\partial z_i^{k_i}} f({\bf z}) = 0,
\end{equation}
where the $P_{i,j}({\bf z})$ are polynomials and for each $i=1,\cdots,n$,
$P_{i,k_i}({\bf z})$ is not the null polynomial, see {\em e.g.} Proposition 2.2
in \cite{lips}. Thus the expression for the susceptibility was shown to be in a
different---and less tractable---class of function than other known
properties of the Ising model. The evidence for this was based on the
observation (not proved) that the anisotropic susceptibility $\chi(v,v'),$
as a function of $v$ with $v'$ fixed has a natural boundary
on the unit circle $|v| = 1.$

For the {\em isotropic} susceptibility, another of us \cite{nica, nicb} 
provided strong confirmation (though again, not a proof)
of this observation
by showing that the circle $|s|=1$ in the complex $s=\sinh{2K}$
plane is a natural boundary. 
These two observations are discussed further in section~\ref{sect:circle},
following a discussion of the general anisotropic case in section~\ref{sect:si}.

In section~\ref{sect:Dfinite} we also prove the important result that while
$\chi(k)$ is not D-finite, $\ch{j}(k)$ {\em is} D-finite for all $j.$

Two other directions in which we have achieved 
substantial progress 
are in the generation of series coefficients for the individual series
$\ch{j},$ continuing work initiated in \cite{nica, nicb}, and
even greater progress in
obtaining the coefficients of the total series $\chi$ using 
nonlinear partial difference equations for the correlation
functions~\cite{mw,p,JM}.

In order to generate the series for the total susceptibility 
$\chi_+$ or $\chi_-$ without computing separately
the $j$-particle contributions,
a more efficient method of series
generation is obtained by first returning to the
expression~(\ref{1.8}) of the susceptibility as the 
sum over all lattice separations
of the two-point correlation functions.

In the scaling limit, the two-point functions were found to satisfy
a nonlinear differential equation of Painlev{\'e} type~\cite{wmtb}
which was then solved to give the leading scaling terms
$|\tau|^{-7/4}$ and $|\tau|^{-3/4}$ exactly \cite{wmtb,bmw}.
In 1980, a discrete analogue of this equation
was discovered by McCoy and Wu~\cite{mw} which holds for the two-point
functions of the lattice Ising model at arbitrary temperature,
and which reduces to the
Painlev{\'e} equation in the scaling limit. In the same year, a
simple set of partial difference equations was derived by Perk~\cite{p},
which reproduced the equation of McCoy and Wu and provided an
additional equation.
Also in the same year, an unrelated set of difference equations, which
can be used to compute the correlation functions on the diagonal, $M=N$,
was obtained by Jimbo and Miwa~\cite{JM}.
Many of these developments are described in some detail in \cite{id}.

The difference equations are valid for arbitrary temperature and
were used by Kong, {\em et al.}~\cite{kapa} to obtain exactly the leading
``short distance" constant terms in the susceptibility both at the
ferromagnetic and anti-ferromagnetic points at $T=\pm T_c.$ This work was
later extended to give the amplitudes of the term $\tau\log{|\tau|}$
\cite{kapb,Kong}. Here we dramatically extend that work by obtaining
all terms in the ``short-distance" part\footnote{When we speak of
the ``short-distance" part, we
include the analytic background term, often ignored in scaling
discussions of the critical region.} $ B_{\rm f/af}$ (see (\ref{6e})
and (\ref{9b})) of the susceptibility to
O$(\tau^{14}).$ Details necessary for the generation of $C(M,N)$ appear
in section~\ref{sect:difference} while in section~\ref{sect:sdc}
we describe the numerical analysis of the $C(M,N)$ that leads us to
conclude that the ``short-distance" terms have the form
\begin{equation}\label{6f}
B_{\rm f/af} = \sum_{q=0}^{\infty} \sum_{p=0}^{\lfloor\sqrt{q}\rfloor}
b^{(p,q)} _{\rm f/af} \tau^q (\log|\tau|)^p \text{ (isotropic)}.
\end{equation}
Although clearly nonanalytic at $\tau=0$ we have denoted these
``short-distance" terms in (\ref{6f})
by $B,$ as a further reminder that they also include the analytical
``background." 
The actual coefficients in (\ref{6f}) can be found in the Appendix.

The quadratic difference equations of Perk \cite{p} can also be used
to generate high- and low-temperature series for $\chi$ and as shown in
section \ref{sect:difference} the series coefficients can be obtained 
in polynomial time!

While some people have expressed 
the view that a polynomial time algorithm for the computation of
the series coefficients 
constitutes a solution, it is clearly preferable to have a closed
form expression. Nevertheless,
a polynomial time algorithm is equivalent to a complete solution
if one seeks only the series coefficients, as
to expand any closed form expression also takes polynomial time. As
the history of the development of these key nonlinear recurrences
described above shows, 
the ingredients for such an algorithm have existed unexploited in the 
literature for many years. 

Analysis of the resulting series of hitherto unimaginable length
combined with the ``short-distance" knowledge contained in (\ref{6f})
leads to a solidly based conjecture specifying completely the
remaining ``scaling" part of the susceptibility $\chi$ of the isotropic
Ising model. Near the anti-ferromagnetic point the ``short-distance"
terms are complete and we simply have for $T>T_c$

\begin{equation}\label{6e}
\beta^{-1}\chi_{\rm af} = B_{\rm af} \text{ (isotropic)}.
\end{equation}
Near the ferromagnetic point we conjecture (for $T>T_c$ or $T<T_c$)

\begin{equation}\label{9b}
\beta^{-1}\chi_{\pm} = C_{0_\pm} (2K_c\sqrt{2})^{7/4}
 |\tau|^{-7/4} F_{\pm} + B_{\rm f} \text{ (isotropic)}
\end{equation}
where the scaling-amplitude functions $F_{\pm}$ 
have (possibly asymptotic) expansions in integer powers of $\tau$
without any of the powers of $\log|\tau|$ present in $B_{\rm f/af}.$
The leading terms are
\begin{equation}\label{9d}
F_{\pm} = 1 +\tau/2 +5\tau^2/8 +3\tau^3/16 -23\tau^4/384 -35\tau^5/768 +
f_{\pm}^{(6)}\tau^6 + {\rm O}(\tau^7)  \text{ (isotropic)}
\end{equation}
where $f_{+}^{(6)} \ne f_{-}^{(6)}.$ In fact the breakdown in equality
is dramatic, and we estimate that $f_{+}^{(6)} = -0.1329693327\ldots,$
and $ f_{-}^{(6)}=-6.330746944\ldots,$ where more accurate values of these
and further terms in~(\ref{9d}) through order
$\tau^{15}$ are given in the Appendix.
In section \ref{sect:anal} we describe the ``short-distance" subtraction 
and analysis on which
the assumed form of the expansion of the scaling-amplitude function 
$F_{\pm}$~(\ref{9d}) is based, while the analysis
leading to the numerical values of the coefficients in~(\ref{9d}) is
sketched in section \ref{sect:ca}.

Aharony and Fisher \cite{AFb,AFa} have predicted a scaling-amplitude function
$F({\rm A\&F})$ that is equal above and below $T_c$ on the assumption that the
Ising model critical region can be described entirely by two 
nonlinear scaling fields. Our exact result (\ref{9d}) is clearly different
and furthermore the explicit expansion (cf. eqs.~(22-24) in \cite{nicb})

\begin{equation}\label{9c}
F({\rm A\&F})= 1 +\tau/2 +5\tau^2/8 +3\tau^3/16 -11\tau^4/192 -
17\tau^5/384 +97\tau^6/3072 + {\rm O}(\tau^7)
\end{equation}
differs from (\ref{9d}) at order $\tau^4.$ This is unequivocal evidence for
the presence of at least one, and almost surely two, irrelevant
operators\footnote{Note that $\tau$ changes sign in~(\ref{9d})
and~(\ref{9c}) as we change from $T > T_c$ to $T < T_c.$}.
There is further possible evidence for irrelevant operators in the
``short-distance" terms (\ref{6f}) which contain powers of $\log{|\tau|}$
beyond the first starting at $\tau^4(\log|\tau|)^2$,
and thus are not of the ``energy" form given by the 
nonlinear field analysis.

An important numerical study investigating corrections to scaling
was that of Gartenhaus and McCullough \cite{GM} who confirmed the $F({\rm
A\&F})$ form in (\ref{9c}) through O$(\tau^3)$ and provided a good estimate of
the term linear in $\tau$ in $B_{\rm f}$ in (\ref{9b}). Estimates of terms to
O$(\tau^2 \log|\tau|)$ in $B_{\rm af}$
in (\ref{6e}) were obtained in \cite{BG}. An attempt \cite{nicb} to
go beyond this using longer series than available in \cite{GM}
was inconclusive other than to indicate the necessity of terms beyond
that predicted by Aharony and Fisher \cite{AFa,AFb}.

Our numerical work began in part as a modest attempt to improve
on \cite{nicb} but expanded to where it now clearly quantifies the effect of
nonlinear scaling fields \cite{AFa} as well as irrelevant operators.
As anisotropy is a marginal operator, extending
the present calculation to the anisotropic square lattice
would, we expect, be extremely helpful in better understanding the effect 
of the irrelevant operator(s)
we have identified. Both this and a study of the susceptibility on the
triangular and hexagonal lattices are projects we hope to tackle in the
near future.

The layout of the paper is as follows.
In section~\ref{sect:defs} we
define the model and give some useful parameterizations.
In section~\ref{sect:integrals} we show how the key integrals
referred to above may be simplified, and provide a short proof of
the assertion that $\ch{j}$, so defined,
is D-finite, even though, as we have seen, $\chi$ is presumably not.
In section~\ref{sect:difference} the computation of the susceptibility series
from the correlation functions by means of nonlinear partial difference
equations is shown to be achievable in polynomial time.
In section~\ref{sect:qseries}
we first discuss the isotropic series in $k$, and then the $q$ series
form of the susceptibility. In this subsection some regularity features
of the coefficients are discovered and the consequences partially developed.
In section~\ref{sect:numerical} we summarize our numerical
work, state our conjecture giving the complete analytic structure
of the isotropic susceptibility, and quantify the effect of irrelevant 
variables in the scaling fields.
Finally, in section~\ref{sect:scaling} we review scaling theory as it
applies to the two-dimensional Ising model, and comment on the relevance of
our results to this theory and to the renormalization group.
The high- and low-temperature series for $\chi_{\rm iso}$ are
given on the WWW at site www.ms.unimelb.edu.au/$\sim$tonyg.

\section{Definitions and notation}\label{sect:defs}

In this section of the paper it will be convenient to formulate the
general anisotropic model. Our later numerical work is confined
to the isotropic model.

The complementary modulus of the elliptic modulus $k$ (\ref{mod})
is given by $k'=\sqrt{1-k^2}$. Here the
prime is not related to the anisotropy. The moduli, $k$ and $k'$,
are related to the elliptic nome, $q$, by
\begin{equation}
k=\left(\frac{\theta_2(0,q)}{\theta_3(0,q)}\right)^2,\qquad
k'=\left(\frac{\theta_4(0,q)}{\theta_3(0,q)}\right)^2.
\end{equation}
The connection between the two moduli yields a well-known theta function
identity. (See section~13.20 of Bateman~\cite{bate} for these and other
formulas.)

In terms of these variables the magnetization takes a particularly
simple form \cite{bsw}
\begin{equation}
{\mathcal M}=(1-(ss')^{-2})^{\frac{1}{8}}=k^{'\frac{1}{4}}
 = \prod_{n=1}^\infty \frac{1-q^{2n-1}}{1+q^{2n-1}}
\end{equation}
for $T < T_c,$
and $0$ otherwise.

The theta functions introduced above
have useful infinite sum and product forms,
which we will subsequently use.
For zero argument these are
\begin{align}\label{2a}
\theta_2(0,q)&=2\sum_{n=0}^\infty q^{(n+1/2)^2}=
2q^{1/4}\prod_{n=1}^\infty(1-q^{2n})(1+q^{2n})^2, \\
\theta_3(0,q)&=1+2\sum_{n=1}^\infty q^{n^2}=
\prod_{n=1}^\infty(1-q^{2n})(1+q^{2n-1})^2, \\
\theta_4(0,q)&=1+2\sum_{n=1}^\infty (-1)^n q^{n^2}=
\prod_{n=1}^\infty(1-q^{2n})(1-q^{2n-1})^2.
\end{align}
Thus,
\begin{equation}\label{2b}
k^{1/4}=\sqrt{2}q^{1/8}\prod_{n=1}^\infty\frac{1+q^{2n}}{1+q^{2n-1}},\qquad
{k'}^{1/4}=\prod_{n=1}^\infty\frac{1-q^{2n-1}}{1+q^{2n-1}}.
\end{equation}

In the elliptic parameterization, one variable is taken to be either $k$ or
$q$. We take the other variable to be the anisotropy parameter.
Onsager~\cite{onsa} used the variable $a$ defined by
\begin{equation}
\sn ia=\begin{cases}is^* & \text{for $T>T_c$,} \\
is' &\text{for $T<T_c$.}\end{cases}
\end{equation}
We will also need the related variable
\begin{equation}
\sn ia'=\begin{cases}i/s' & \text{for $T>T_c$,} \\
i/s^* &\text{for $T<T_c$.}\end{cases}
\end{equation}
In some contexts, it will be useful, following ref.~\cite{kapa}, to use
instead the variables $\alpha$ and $\alpha'$ defined by
\begin{align}\label{alpha}
\cot\alpha&=-i\sqrt{k}\sn ia=\sqrt{s's^*}=\sqrt{s'/s}, \\
\cot\alpha'&=-i\sqrt{k}\sn ia'=\sqrt{1/(s's^*)}=\sqrt{s/s'}.
\end{align}
Obviously $\cot\alpha=\tan\alpha'$.
The isotropic values of these variables are $\sn ia=\sn ia'=i/\sqrt{k}$ and
$\alpha=\alpha'=\pi/4$.

We observe that $q^{1/2}=x$ where $x$ is the variable of
ref.~\cite{bsw}, defined by
\begin{equation}
e^{-2K}=x^{1/2}\prod_{n=1}^\infty\frac{(1-x^{8n-7})(1-x^{8n-1})}
{(1-x^{8n-5})(1-x^{8n-3})},
\end{equation}
in which the spontaneous magnetization has a simple product form.
The elliptic modulus of ref.~\cite{bsw}
is related to our $k$ by a Landen transformation~\cite[section 15.6]{baxbook}.

In reference~\cite{bsw} it was also noted that
the first terms of the isotropic ($K=K'$) high-temperature susceptibility
have a simple product expression, which breaks down at order $q^{8/4}$.
This is explained by the expansion of the high-temperature
susceptibility in multi-particle states described in the Introduction.
The contribution of one-particle states has a product form, and the
first three-particle state contributes at order $q^{8/4}$. This one-particle
product form is
\begin{equation}
\chiso{1}=q^{1/4}\prod_{n=1}^\infty \frac{(1+q^{n/4})^2 (1-q^{4n})^2}
{(1-q^{n/4})^2(1+q^n)^2}=\frac{q^{1/4}}{\theta_3(0,q)}\prod_{n=1}^\infty
\frac{(1+q^{n/4})^2 (1-q^{2n})^3}{(1-q^{n/4})^2}.
\end{equation}
(Strictly speaking the product form above is not the product
form of ref.~\cite{bsw} which actually breaks down only
at order $q^{9/4}$. They differ by a factor which first contributes
at order $q^{8/4}$ and the absence of this factor in ref.~\cite{bsw}
exactly compensates
for the addition of the first term of $\chiso{3}$.)

Attempts to do the same for the low-temperature series have failed as
we do not know of a product form for $\chiso{2}$. The function {\rm K} has
one, namely
\begin{equation}
{\rm K} = \half\pi \theta_3^2(0,q),
\end{equation}
while the function {\rm E} is given in terms of $q$ by the formula \cite{bate}
\begin{equation}
{\rm E}=\frac{\theta_3^4(0,q)+\theta_4^4(0,q)}{3\theta_3^4(0,q)}{\rm K}-
\frac{1}{12{\rm K}}\cdot\frac{\theta'''_1(0,q)}{\theta'_1(0,q)}.
\end{equation}

\section{Integral formulae for $\ch{j}$}\label{sect:integrals}
In this section we present integral expressions for the $\ch{j}$ and
show that they define D-finite functions.

\subsection{Trigonometric form of integrals}\label{sect:si}
We will start with the form given by Yamada \cite{yamb, yamd, yama, yamc, yame}
for the $j$-particle
contribution $\ch{j}$ in terms of elliptic variables,
and derive from it an expression in terms of trigonometric/hyperbolic
variables. The integral expression is
\begin{equation}\label{3.1}
\ch{j}=\frac{k^{j/2}}{(2\pi)^j j!}\int_0^{4{\rm K}}du_1 \cdots
\int_0^{4{\rm K}} du_j \left(G^{(j)}\right)^2
\frac{1+\prod_{n=1}^j x_n}{1-\prod_{n=1}^j x_n}\cdot
\frac{1+\prod_{n=1}^j z_n}{1-\prod_{n=1}^j z_n},
\end{equation}
where the modulus of the complete elliptic integral $\rm K$ is $k$, and
this same modulus is assumed in all Jacobi elliptic functions which
appear below.
Considered as a function of one of the $u_n$, the integrand has a single
simple pole on the real axis which derives from
the last factor in the integrand. The contour of integration is deformed
in the vicinity of the pole so that only half the residue is taken. 
The function $G^{(j)}$ can be written as the product
\begin{equation}
G^{(j)}=\prod_{1\le m<n\le j} h_{mn},
\end{equation}
and $z_n$, $x_n$ and $h_{mn}$ are given by
\begin{align}\label{hij}
z_n=e^{i\omega_n}&=\frac{\sn\half(u_n+ia')}{\sn\half(u_n-ia')}
=-\frac{\sn ia'\cn u_n+\cn ia'\sn u_n}{\sn ia'\dn u_n-\dn ia'\sn u_n},
 \nonumber \\
x_n=e^{-\gamma_n}&=k\sn\half(u_n+ia')\sn\half(u_n-ia')
=k\frac{\cn ia'-\cn u_n}{\dn ia'+\dn u_n}, \nonumber \\
h_{mn}&=-\sqrt{k}\sn\half(u_m-u_n).
\end{align}
Note that the trigonometric/hyperbolic variables, $\omega_n$ and $\gamma_n$
are related to the elliptic variables, $u_n$ and $a'$ by
the functional equation
\begin{equation}
\exp(\pm\half i\omega_n-\half\gamma_n)= \sqrt{k}\sn\half(u_n\pm a').
\end{equation}
The choice of $a'$ rather than $a$ as anisotropy parameter is arbitrary
because of the symmetry under interchange of horizontal and vertical lattice
axes. Choosing $a'$ at this point results in
expressions for $\omega_n$ and $\gamma_n$ which are
equivalent to those of Onsager.

The mapping between the elliptic parameterization and the
trigonometric/hyperbolic one
was described in ref.~\cite{nicb} for the isotropic case. The formulas
below are a generalization of this mapping.
It is simpler to use the trigonometric parameterization
than it is to use the elliptic parameterization for
numerical series generation.

The variables $\omega_n$ and $\gamma_n$ defined above satisfy the identities
given in Appendix~2 of Onsager's paper~\cite{onsa},
\begin{align}\label{gamma}
\cosh\gamma_n&=(\cn ia\dn u_n-k\dn ia\cn u_n)/M_n \\
\sinh\gamma_n&=-i{k'}^2\sn ia/M_n \\
-\cos\omega_n&=(\dn ia\cn u_n-k\cn ia\dn u_n)/M_n \\
\sin\omega_n&={k'}^2\sn u_n/M_n,
\end{align}
with
\begin{equation}
M_n=\dn ia\dn u_n-k\cn ia\cn u_n.
\end{equation}
Notice that it is $a$ which appears in these formulas rather than $a'$.
The mapping from the variables
$u_n$ and $a$ to the variables $\omega_n$ and $\gamma_n$ is conformal, as is
seen from the formulas, also given by Onsager
\begin{align}
-\partial\omega_n/\partial u_n&=\partial\gamma_n/\partial a={k'}^2/M_n
=i\sinh\gamma_n/\sn ia \nonumber\\
\partial\gamma_n/\partial u_n&=
\partial\omega_n/\partial a={k'}^2k\sn ia\sn u_n/M_n.
\label{conformal}
\end{align}
Finally, Onsager gives the functional equation
\begin{equation}
\cot\half(\omega_n-i\gamma_n)=(1+k)\jacobisc\half(u_n+ia)\nd\half(u_n+ia).
\end{equation}

To make the change of variables, we first note that $\phi_n=\pi$
corresponds to $u_n=0$, and $\phi_n=0$ corresponds to $u_n=2{\rm K}$.
Using eq.~(\ref{conformal}), we see that
\begin{equation}
\int_{-\pi}^\pi d\omega\,\cot\alpha/\sinh\gamma\ldots =
\sqrt{k} \int_0^{4{\rm K}}du\ldots,
\end{equation}
which implies
\begin{equation}
\ch{j}=\frac{\cot^j\alpha}{j!}\int_{-\pi}^{\pi}\frac{d\omega_1}{2\pi} \cdots
\int_{-\pi}^{\pi} \frac{d\omega_{j-1}}{2\pi} \left(\prod_{n=1}^j \frac{1}
{\sinh\gamma_n}\right)
\left(\prod_{1\le i<k\le j}h_{ik}\right)^2
\frac{1+\prod_{n=1}^j x_n}{1-\prod_{n=1}^j x_n}.
\label{chij}
\end{equation}\label{3.13}
The condition $\omega_1+\ldots+\omega_j=0$ mod $2\pi$, which results
from having performed one of the integrations, is assumed. In terms
of $\omega_n$, $k$ and $\alpha$, the quantities $x_n$ and $\sinh\gamma_n$ can
be expressed as
\begin{align}\label{3.14}
x_n&=\cot^2\alpha\left[\xi-\cos\omega_n-
\sqrt{(\xi-\cos\omega_n)^2-(\cot\alpha)^{-4}}\right], \\
\sinh\gamma_n&=
\cot^2\alpha\sqrt{(\xi-\cos\omega_n)^2-(\cot\alpha)^{-4}}
\end{align}
with
\begin{equation}\label{3.16}
\xi=\left(1+1/(k\cot^2\alpha)\right)^{1/2}
\left(1+k/\cot^2\alpha\right)^{1/2}=\left(1+(s')^{-2}\right)^{1/2}
\left(1+(s^*)^{-2}\right)^{1/2}.
\end{equation}
In the isotropic case, these reduce to
\begin{align}
x_n&=s+s^{-1}-\cos\omega_n - \sqrt{(s+s^{-1}-\cos\omega_n)^2-1}, \\
\sinh\gamma_n&=\sqrt{(s+s^{-1}-\cos\omega_n)^2-1}.
\end{align}
Here $s=\sqrt{k}$ for high-temperature and $s=1/\sqrt{k}$ for low
temperature, although the distinction is irrelevant as the
dependence of the integrand on $s$ and $1/s$ is
symmetric.
Finally
\begin{align}
h_{ik}&=\cot\alpha\frac{\sin\half(\omega_i-\omega_k)}
{\sinh\half(\gamma_i+\gamma_k)}=\frac{1}{\cot\alpha}
\frac{\sinh\half(\gamma_i-\gamma_k)}{\sin\half(\omega_i+\omega_k)} \\
&=\frac{2(x_i x_k)^{1/2}\cot\alpha\sin\half(\omega_i-\omega_k)}{1-x_i x_k}.
\end{align}

When $j=1$ or $j=2$, the integrals can be rewritten in terms of known
functions as was noted in the introduction for the isotropic case.
When $j=3$ Glasser~\cite{LG98} showed that the integrals can be written as an
integral involving the square root of a polynomial of degree higher
than 4. Such integrals are often called 
{\em hyperelliptic} integrals, and are a special case of
{\em Abelian} integrals.

The corresponding anisotropic expressions are
\begin{align}
\ch{1}=&\frac{1}{(k^{1/2}-k^{-1/2})^2}\left[2\csc 2\alpha+
\sqrt{(k^{1/2}-k^{-1/2})^2+4\csc^2 2\alpha}\right],\\
\ch{2}=&\frac{k^{1/2}}{1+k}\sqrt{(k^{1/2}-k^{-1/2})^2+4\csc^2 2\alpha}
\,\chiso{2}.
\end{align}

As we show in section~\ref{sect:qseries}, we can say more
about the general form of the expansion,
based on inspection of the long series published by Nickel~\cite{nica,nicb},
as well as more recent extensions reported here, giving hope that
there is still more regularity to be found.

\subsection{Natural boundaries}\label{sect:circle}

As mentioned in the introduction, there were observations, first
in~\cite{GE96} and then in~\cite{nica,nicb}, that strongly suggested the
susceptibility of the Ising model is a function with a natural
boundary unlike the free-energy or magnetization.  We
expand and clarify those observations here in the light of our new
knowledge of the general anisotropic case discussed in section~\ref{sect:si}
and the additional numerical work on the isotropic limit described in
section~\ref{sect:numerical}.

First, we expect that as described in~\cite{nica} , the $\cu{j}$ given by
the integrals (\ref{3.13}) multiplied by the factors outside the
summation in eq.~(\ref{1a}) or~(\ref{1b}) will be singular at the
symmetry points of the
integrand and where the denominator factor $1-\prod_{n=1}^j x_n$
vanishes. The symmetry point condition requires all $\omega_n$ to be equal
and given by $\omega_n = \omega = 2\pi m'/j,$ $ m'=1,2,\ldots j.$ The
vanishing of the denominator factor requires the $x_n$, now all equal, to be
given by $x_n = x = \exp(2\pi i m/j),$ $ m=1,2,\ldots j.$ Equivalently, from
the explicit formula (\ref{3.14}),
\begin{equation}\label{last}
\cot^2(\alpha)(\xi-\cos(2\pi m'/j)) = \cos(2\pi m/j). 
\end{equation}
With $\cot^2(\alpha) = s'/s$ from (\ref{alpha}) and $\xi$ given by (\ref{3.16}),
we find (\ref{last}) can be reduced to

\begin{equation}\label{3.24}
\cosh(2K)\cosh(2K') - \sinh(2K)\cos{\frac{2\pi m}{j}}
-\sinh(2K')\cos{\frac{2\pi m'}{j}}=0
\end{equation}
with $m,m'=1,2,\ldots j$ as discussed above. It will be noted that the
left-hand side of (\ref{3.24}) is the denominator in the Onsager integral
for the free-energy and thus we find the (to us) surprising result
that the singularity of $\cu{j}$, a property of the Ising model in a
magnetic field, is intimately connected with a property in zero
field.

The full $\chi$, being a sum of $\cu{j}$, will naively be
expected to be singular at a dense set of points and thus have the
Onsager line (\ref{3.24}) as a natural boundary.
The presence of natural boundaries has implications for expansions
about the physical singularity points $s = \pm 1$ that are
necessary to understand corrections to scaling. We briefly explore 
some of these implications but restrict ourselves for simplicity
to the ferromagnetic point $s = 1$ in the isotropic model. We
also make the plausible assumption that the singularity in each
$\cu{j}$ closest to $\tau=0$ is the most important for determining,
in expansions of $\chi$, the $\tau^p$ large $p$ asymptotics and this
considerably simplifies the discussion\footnote{Note that singularities
 on $|s|=1$ map to points on the imaginary axis in the complex $\tau$ plane.}.

Let $\tau_j = i\sin{\theta_j}$ be that singularity in $\cu{j}$ that
is closest to the ferromagnetic $\tau = 0$; $\theta_j$ is fixed by
$\cos{\theta_j} = (1 + \cos{\phi_j})/2$ with $\phi_j = 2\pi/j.$ Choose
the branch-cut arising from this singularity to lie along the 
imaginary $\tau$ axis and directed away from $\tau = 0.$ Take 
$\tau = i\mathcal{T}$ to be a point on the branch-cut. Provided the
(positive) deviation
\begin{equation}\label{new26}
\delta\theta_j = \arcsin{\mathcal{T}} - \theta_j
\end{equation}
is not too large, the discontinuity across the cut can be estimated
from the linearized singularity equations (14) in \cite{nica} and (12)
in \cite{nicb}.
For $j > 2$ a more general result that includes the first order
correction is
\begin{eqnarray}\label{new27}
{\rm Disc}(\cu{j}) &=& -C_j i^{j^2}/\sin^2{\phi_j}
[\delta\theta_j\sin{\theta_j}/
\sin^2{\phi_j}]^{(j^2-3)/2} \nonumber \\
&\times&\left[1 + \delta\theta_j\sin{\theta_j}
\left\{\frac{\cos{\theta_j}}{4\sin^2\theta_j}
 + \frac{j^2-4}{8\cos{\theta_j}(1+\cos{\theta_j})} - \frac{1}{2}\right\} + 
{\rm O}(\delta\theta_j^2)\right]
\end{eqnarray}
where
\begin{equation}\label{new28}
C_j = 2\sqrt{2}(2/\pi)^{(j-3)/2}(j/2)^{(j^2-4)/2}(\Pi_{m=1}^j \Gamma(m))
/\Gamma((j^2-1)/2) \approx 3.7655 j^{-1/12} 2^j \exp(-j^2/4).
\end{equation}
The last equality is valid only in the large $j$ limit.
To obtain the discontinuity in say $\chi_+$, we must first sum the
discontinuities in $\cu{j}$, $j$ odd, and this we can crudely
estimate by integrating over $j$, keeping only the leading exponential
factors in (\ref{new26}) and (\ref{new27}) and making a small angle 
approximation
as well. Essentially the same formula is obtained for the
discontinuity in $\chi_-$ on summing over $j$ even; in either case we find
\begin{equation}\label{new29}
{\rm Disc}(\sum_{j} \cu{j})
\sim \int_{\sqrt{2}\pi/\mathcal{T}}^{2\pi/\mathcal{T}}
dj [(\mathcal{T}j/(\sqrt{2}\pi)-1)/(2\sqrt{\rm e})]^{j^2/2}
\end{equation}
where the sum is over $j$ odd or $j$ even, and the
lower limit is simply the restriction to those $j$ values that
 contribute, while the upper limit roughly defines the limit of validity of
the linearized approximation. The precise value of this limit is not
important since the integrand has a maximum well below the limit. For 
large $j$ only the maximum of the integrand matters and (\ref{new29})
 reduces to
\begin{equation}\label{new30}
{\rm Disc}(\chi) \sim \exp(-39.76/\mathcal{T}^2)
\end{equation}
in which the numerical coefficient of $1/\mathcal{T}^2$ is $\pi^2x^3/(2(x-1))$
with $x$ the smallest real solution of $2\log((x-1)/2) + 1/(x-1) = 0.$
Keeping terms such as the correction term in (\ref{new27})
and the $2^j$ in (\ref{new28}) in the steepest descent analysis lead
to O$(1/\mathcal{T})$ corrections to the exponent in (\ref{new30}) so
we conjecture that the right hand side of (\ref{new30}) is exactly the
leading exponential. It is interesting that the discontinuity (\ref{new30})
is similar to that found in weak coupling field theory expansions, but the 
mechanism producing the cut here does not seem to be related in any way
to instantons.

The additional singularities that $\chi$ has at $\tau=0,$ namely
terms such as the divergent ``scaling" $|\tau|^{-7/4}$ or the ``short-
distance" powers of $\log|\tau|,$ are a complication we do not know how
to handle in any rigorous fashion\footnote{ If we knew that the
``short-distance" corrections $B_{\rm f}$ in (\ref{9b})
formed a convergent sequence, then a subtraction process similar to
that described in section \ref{sect:anal} could be carried out here. The
numerical evidence from section \ref{sect:anal}
 is that the point $\tau=0$ in the
scaled and pole subtracted $\chi_{\pm}$ is no longer a branch point
singularity and hence the final result (\ref{eqcut}) is justifiable and would
apply directly to the scaling-amplitude function 
$F_{\pm}.$ Unfortunately, we do not have
any independent information to suggest
that the ``short-distance" sum (\ref{6f}) is
convergent and thus the argument leading to (\ref{eqcut}) is at best
suggestive that one or the other (or both) of the ``short-distance" or
``scaling" sequences are asymptotic.}.
As a consequence we will simply
ignore them and make the simplest, yet reasonable, assumption that
they make an additive contribution not relevant for understanding the
effect of the natural boundary. 
Then the cut discontinuity (\ref{new30}) would imply a divergent
behavior in the $\tau$ expansion of $\chi$. That is to say,
the coefficient of $\tau^p$ in the limit $p \rightarrow \infty$ will
diverge as $\Gamma(p/2)/a^{p/2}$ with $a \approx 39.76$.
This follows from a contour integral around the origin distorted to 
run on either side of the cut imaginary axis. The contribution of the 
cut discontinuity to the coefficient, $C_p$, of $\tau^p$ in the 
expansion is then 
\begin{equation}\label{eqcut}
C_p \propto \int d\mathcal{T}/\mathcal{T}^{p+1} \exp(-a/\mathcal{T}^2) \sim
 \Gamma(p/2)/a^{p/2}.
\end{equation}

 We know little about the cut discontinuity on the circle $|s|=1$
other than what we have deduced near $s=1$ as given by (\ref{new30}). 
However, the fact that the amplitudes of the singularities of
$\cu{j}$ on $|s|=1$ vary dramatically with order $j$ almost certainly
implies there will be no cancellation of singularities in the sum of 
$\cu{j}$ that defines $\chi.$ Furthermore, the variation with order
means that there is no length scale at which, as one approaches the
circle of singularities $|s|=1,$ $\chi_{\rm iso}$ can be smooth. It
is these two points taken together that we consider overwhelming
evidence that, in the isotropic case at least, $\chi$ has a natural
boundary that is the entire $|s|=1$ circle.
We do not imply by this that all points on the circle are
equally ``singular". As argued above, the existence of an asymptotic
expansion about $s=1$ seems likely. A very different situation arises
at a point such as $s=i.$ While the singularity in $\cu{j}$ on the
circle $s=\exp(i \theta)$ nearest $s=1$ lies at a distance $\Delta \theta
= {\rm O}(1/j)$ for large $j,$ the corresponding nearest distance from $s=i$
is $\Delta \theta = {\rm O}(1/j^2).$ Furthermore this latter singularity is
larger in its leading amplitude than the former by a factor roughly
$j^{(j^2)/2}.$ The reduction in distance and dramatic increase in
amplitude suggests that an asymptotic expansion about $s=i$ is not possible,
but this has not been proved.

For the anisotropic case we have not analyzed (\ref{3.13}) in detail
so we do not have the necessary amplitude information to make the
same claim directly. However, we can take the extreme anisotropic
limit of $s'$ infinitesimal (but not $0$) and find that the Onsager line
(\ref{3.24}) has now come very close to the circle $|v|=1.$
 At this point we
can connect to the work in \cite{GE96}. There it was observed that if
$\chi(K,K')$ was written as

\begin{equation}\label{3.25}
\chi(v,v') = \sum H_n(v) {v'}^n 
\end{equation}
the $H_n$ would be singular at a dense set of points on $|v|=1$ as
$ n \rightarrow \infty.$ Furthermore, if as in the discussion above, $v'$
 is infinitesimal (but not $0$) the amplitudes of the singularities vary
dramatically with order $n$ and the same conclusion as in the isotropic
case is reached. Given that the Onsager line (\ref{3.24}) is a natural
boundary in two extremes, it seems highly probable that it is also a
natural boundary at all intermediate anisotropy values.

We conclude this section by contrasting the above behavior
of the susceptibility with that of the
free-energy and magnetization which are only singular at an isolated
set of points, not a dense set. This is precisely what one expects
for a D-finite function, as in that case we have the following\footnote{Due
to M. Bousquet-M\'elou, private communication.}
\newtheorem{mireille}{Theorem}
\begin{mireille}
Let $f(x,y) = \sum_{n\ge 0} y^n H_n(x)$ be a D-finite
series in $y$ with rational coefficients. For $n \ge 0,$ let $S_n$ be
the set of poles of $H_n(x)$; let $S=\bigcup_n S_n.$ Then $S$ has only
a finite number of accumulation points.
\end{mireille}
This is observed in practice. The anisotropic 
magnetization and free-energy each have 
exactly one accumulation point~\cite{Gu00}, while the (non-D-finite)
susceptibility appears to have an infinite number~\cite{GE96}.

\subsection{$\cu{j}$ is D-finite}~\label{sect:Dfinite}
This remarkable result, that 
$\cu{j}$ (or equivalently $\ch{j}$) is D-finite while $\chi$ is not,
follows from the results
of Lipshitz~\cite{lips} (see also Zeilberger~\cite{z}),
who gives several basic definitions
and theorems concerning D-finiteness of series in several variables.
The integrand of the trigonometric/hyperbolic form of $\ch{j}$ is an
algebraic function of the variables $c_j=\cos\omega_j$ and $k^{1/2}$,
and is thus
a D-finite function in these same variables. Then by Theorem~2.7 of
ref.~\cite{lips}, integrating over one or more variables preserves
D-finiteness which implies the result. The term D-finite is synonymous
with {\em holonomic} in much of the literature. 

Kashiwara and Kawai~\cite{Kash} have shown that any Feynman
diagram is holonomic, whereas an infinite sum of such diagrams
may not be. This is just the phenomenon
we observe here. At first glance 
it appears that Kashiwara's definition of holonomic
differs from that used here, but this is not so. The point is that
the definition of D-finite functions of more than one variable
requires that the underlying system of partial differential equations
be such that only a finite number of initial conditions are needed to
specify the function. Such systems are called ``maximally over-determined"
or ``holonomic" in the analysis literature. In the single variable case,
the question of a finite number of initial conditions is clearly automatically
satisfied.

Motivated by the above observation, we have attempted to find
linear differential equations with polynomial coefficients in $k^{1/2}$,
or equivalently, linear recurrences for the series coefficients, of
$\chiso{3}$ and $\chiso{4}$. With the available series of order
$k^{257/2}$ and $k^{62}$ respectively, calculated using the methods 
of numerical integration described in \cite{nica,nicb}, we have
ruled out any such recurrences of depth 14 with coefficients of degree
15 for $\chiso{3}$, and of depth 6 with coefficients of degree 7 for
$\chiso{4}$.
Thus while these functions are provably D-finite, it is clear that
the generating differential equation will be a fairly cumbersome
object.

We have also attempted to fit these series as polynomials in the
complete elliptic integrals K and E with polynomial coefficients
in $k$ and obtained similar negative results. 

\section{Correlation functions and difference equations}\label{sect:difference}

In this section we give details of our more efficient method of series
generation based on summing the correlation functions, obtained
by means of nonlinear
recurrences, as outlined in the Introduction.

The equations we used for generation of the very long series
are the ones given by Perk~\cite{p}. Here we present a slight generalization
due to McCoy and Wu~\cite{mw2} which keeps track of the separate
multi-particle components. The expansions of
the two-point correlation functions in multi-particle components are
analogous to the corresponding expansions~(\ref{1a}) and~(\ref{1b})
for the susceptibility
\begin{equation}
C(M,N;\lambda)=\begin{cases} k^{-1/2}(1-k^2)^{1/4}\sum_{n=0}^\infty
\lambda^{2n+1}\Ch{2n+1}(M,N) &
\text{for $T>T_c$} \\
(1-k^2)^{1/4}\sum_{n=0}^\infty\lambda^{2n}\Ch{2n}(M,N) &
\text{for $T<T_c$,} \end{cases}
\end{equation}
with $\Ch{0}(M,N)=1$ and for $j>0$
\begin{align}\label{ci}
&\Ch{j}(M,N)=\\
&\quad\frac{\cot^j\alpha}{j!}\int_{-\pi}^{\pi}\frac{d\omega_1}{2\pi} \cdots
\int_{-\pi}^{\pi} \frac{d\omega_j}{2\pi} \left(\prod_{n=1}^j \frac{1}
{\sinh\gamma_n}\right)
\left(\prod_{1\le i<k\le j}h_{ik}\right)^2 \left(\prod_{n=1}^j x_n\right)^M
\cos(N\sum_{n=1}^j \omega_n).\nonumber
\end{align}
The fugacity $\lambda$ is associated with the number of particles, and
$C(M,N)=C(M,N;1)$.

With these definitions the quadratic partial difference equations are
\begin{align}\label{pd1}
&s^2 [C(M,N;\lambda)^2-C(M,N-1;\lambda)C(M,N+1;\lambda)]\nonumber\\
&\quad+[C^*(M,N;\lambda)^2-C^*(M-1,N;\lambda)C^*(M+1,N;\lambda)]=0\\
\label{pd2}
&{s'}^2 [C(M,N;\lambda)^2-C(M-1,N;\lambda)C(M+1,N;\lambda)]\nonumber\\
&\quad+[C^*(M,N;\lambda)^2-C^*(M,N-1;\lambda)C^*(M,N+1;\lambda)]=0\\
\label{pd3}
&ss'[C(M,N;\lambda)C(M+1,N+1;\lambda)-C(M,N+1;\lambda)C(M+1,N;\lambda)]
 \nonumber
\\
&\quad=C^*(M,N;\lambda)C^*(M+1,N+1;\lambda)-C^*(M,N+1;\lambda)
C^*(M+1,N;\lambda).
\end{align}
The object $C^*(M,N;\lambda)$ is the correlation function on the dual
lattice and is obtained from $C(M,N;\lambda)$
by replacing $s'$ with $s^*$ and $s$ with ${s'}^*$.
Equation~(\ref{pd3}) holds for all $M$ and $N$. When $M=N=0$
equations~(\ref{pd1}) and~(\ref{pd2}) must be replaced by
\begin{align}
&C^*(1,0;1)=\sqrt{1+s^2}-sC(0,1;1)\\
&C^*(0,1;1)=\sqrt{1+{s'}^2}-s'C(1,0;1).
\end{align}
We do not know of $\lambda\ne1$ versions of these equations.

These equations are nearly enough to determine all two-point functions
completely. For the isotropic expansion, all that is lacking is
either the high or the low
temperature set of diagonal correlations ($M=N$). From either one of these
the other can be obtained using equation~(\ref{pd3})
with $M=N$. When $\lambda\ne1$ we have used the integral
formula~(\ref{ci})
to compute the diagonal correlation functions.
For the rest of this section we focus on the case $\lambda=1$ where two
superior methods for obtaining the diagonal correlations are available.
From the purely computational point of view the
difference equations of Jimbo and Miwa~\cite{JM} are almost certainly the
most efficient and to be preferred. However, from the point of view
of understanding the analytical structure of the correlations the
original Toeplitz determinants~\cite{KO,MPW,mwbook} are better. We also find
that for the numerical computations we have carried out so far the
evaluation of the determinants is only a small fraction of the total
project time so efficiency is not yet an issue.

We will restrict ourselves in the following to the isotropic
lattice. In that case and for $N>0,$
the diagonal correlation $C(N,N)$ is the determinant of an $N \times N$ 
Toeplitz matrix with elements
$a_{i,j} = a_{i-j}$ that are the integrals~\cite{mwbook} 

\begin{equation}\label{7a}
a_n = (2\pi)^{-1} \int_0^\pi d\theta (s - \exp{(-2i\theta)}/s)
\exp{(-2i\theta n)}/\sqrt{\tau^2 + \sin^2\theta}.
\end{equation}

These integrals apply both above and below $T_c$ and furthermore
since $\tau \rightarrow -\tau$ corresponds to $ s \rightarrow 1/s, $
one can establish from the
integral in~(\ref{7a}) the relations $a_{-n-1}(\tau)=-a_n(-\tau).$
Explicit formulae
for $a_n$ for small $n$ can be given in terms of elliptic integrals E
and K but these expressions are not particularly enlightening and are not
needed here. Rather we need the series expansions of~(\ref{7a}), either
in $s,$ $1/s$ or $\tau$ depending on the application.

 The series expansions in $s$ and $1/s$ both for
$a_n$ and $C(N,N)$ are completely straightforward with computer packages
such as Maple that automatically handle the multiple precision
arithmetic required. Furthermore, these same packages can be set to
treat the $C(M,N)$ in the recursion
formulae~(\ref{pd1})--(\ref{pd3}) as
series and thus very little programming is necessary to generate the
susceptibility.
Admittedly, some steps
need to be taken to conserve time and/or memory resources but this is
very hardware dependent and will not be described here.
 What is worth noting, however, is that for a series
to order $N$ the recursion formulae require {\rm O}$(N^2)$
multiplications of series of length $N$ and thus in a naive
implementation, {\rm O}$(N^4)$ multiplications. Since the word
length grows linearly with $N$ the algorithm has complexity of at
most {\rm O}$(N^6)$. There are more efficient ways to multiply long
series and numbers with a large number of digits \cite{Knu} but we
have not found it necessary to explore these options.

Timing proportional to $N^6$ is observed in practice, in our
implementation of the recursion in Maple. We have generated high-temperature
series of order 323 and low-temperature series of order 646 for the
isotropic susceptibility. The entire calculation took 123 hours on a
500MHz DEC Alpha with 21164 processor running Maple V version 5.1. We
have also obtained shorter anisotropic series in this way (either the nearest
off-diagonal correlation functions, or additional assumptions are needed).
As lattice anisotropy is a marginal operator, we hope that an
extension of this calculation will be very revealing.

The series in $\tau$ is most easily obtained by expressing the
Toeplitz element integral (\ref{7a}) in terms of hypergeometric
functions. To show this connection we start by writing

\begin{equation}\label{7aa}
\pi a_n/2 = \sqrt{1+\tau^2} A_s - \tau A_c
\end{equation}
in terms of the real integrals

\begin{align}\label{7ab}
A_s =\int_0^{\pi/2} d\theta \sin(\theta) \sin(\nu \theta)
   /\sqrt{\tau^2+\sin^2{\theta}}, \nonumber \\
A_c =\int_0^{\pi/2} d\theta \cos(\theta) \cos(\nu \theta)
   /\sqrt{\tau^2+\sin^2{\theta}},
\end{align}
where $\nu =2n+1.$
The required symmetry $a(-\nu,\tau)=-a(\nu,-\tau)$ is explicit in
eqs.~(\ref{7aa}) and (\ref{7ab}) and by standard integration by parts
manipulation one can show that the cosine integral satisfies the
differential equation
\begin{equation}\label{7ac}
(1+\tau^2)(d/d\tau)\tau(d/d\tau)A_c - \nu^2 \tau A_c =0 
\end{equation}
while the sine integral can be expressed as the derivative
\begin{equation}
A_s = -(\tau/\nu) (d/d\tau)A_c.
\end{equation}
Furthermore, a direct evaluation of the integral in (\ref{7ab}) for
small $\tau$ yields $A_c = -\ell_\nu +{\rm o}(\tau)$ with

\begin{equation}\label{7c}
\ell_\nu = \log(|\tau|/4) + \psi(\nu/2)/2 + \psi(-\nu/2)/2 - \psi(1/2),
\end{equation}
and this initial condition together with (\ref{7ac}) completely
determines $A_c.$ Since (\ref{7ac}) can be recognized as the
hypergeometric differential equation in the variable $z=-\tau^2,$ we
can immediately write the solution as \cite{AS}

\begin{align}
A_c =& -\ell_\nu F(\nu/2,-\nu/2;1;-\tau^2) +
 \sum_{k=1}^\infty (\nu/2)_k (-\nu/2)_k /(k!)^2 (-\tau^2)^k \nonumber \\
&\times (\psi(k+1)-\psi(1)+\psi(\nu/2)/2-\psi(k+\nu/2)/2
 +\psi(-\nu/2)/2-\psi(k-\nu/2)/2).
\end{align}
On Taylor expansion we now obtain

\begin{align}\label{7b}
\pi a_n/2 & = \nu^{-1} + \ell_{\nu}\tau +
 [\nu(\ell_{\nu} - 1/2) + \nu^{-1}]\tau^2/2 +\nu^2
(\ell_{\nu} - 1)\tau^3/4 \\
& \mbox{ } + [\nu^3(\ell_{\nu} - 5/4)-\nu - 2\nu^{-1}]\tau^4/16 +
[\nu^4(\ell_{\nu} - 3/2)-4\nu^2(\ell_{\nu} - 1/2)]\tau^5/64
 + {\rm O}(\tau^6) \nonumber
\end{align}
which can be extended as required.

From the expression~(\ref{7b}) one can conclude that $C(N,N),$ as an
$N \times N$ determinant, will contain logarithmic terms
$\tau^q(\log{|\tau|})^p$
with $q \ge p$ and $p \le N$---barring cancellation.
However, there is cancellation and the key qualitative features of
the cancellation can be deduced simply by using an alternative
representation for the determinant. In particular, by systematically
subtracting rows and columns one can show that an equivalent
determinant has matrix elements $ a_{i,j}$ which are of the same
integral form as the $a_n = a_{i,j}$ in eq.~(\ref{7a}) except for
the replacement of the exponential $\exp(-2i \theta n)$ by the product 
$\exp(-i\theta n) (2 \sin(\theta))^{i+j-2}.$
The new matrix is no
longer of Toeplitz form but for our purposes here it is the better
representation because of the powers of $\sin(\theta)$ which have the
effect of shifting the $\log|\tau|$ singularity of the integrals to
higher order in $\tau.$
Indeed one can show from the new integrals that
the leading singular behavior of each matrix element $a_{i,j}$
is proportional to $\tau^{i+j-1}\log|\tau|$ and this is sufficient to
show that in the logarithmic terms $\tau^q (\log|\tau|)^p$ in $C(N,N)$ one
must have $q \ge p^2.$ We have not attempted to pursue this
argument to deduce $C(N,N)$ analytically but rather have resorted to a
numerical small $N$ evaluation of $C(N,N)$ using~(\ref{7b}) and then fitting
to obtain formulae valid for general $N.$ 
Our results for the
leading logarithm term are summarized
by the expression

\begin{eqnarray}\label{7d}
\sqrt{s} C(N,N,\tau)& =& C(N,N,\tau=0) \sum_{p=0}^\infty 4^p (\log |\tau|
+ \mathcal{L}_N)^p (N\tau/4)^{p^2}\\ \nonumber
& & \{\prod_{k=1}^{p-1}(N^{-2} 
- k^{-2})^{p-k}\}\{1 + (1 + 2(N^2 - p^2))\tau^2/8 + {\rm O}(\tau^3)\}
\end{eqnarray}
in which we have used $\mathcal{L}$
 to denote the discrete logarithm, i.e.
\begin{equation}\label{7e}
\mathcal{L}_N = \psi(N+1)/2 + \psi(N)/2 - \psi(1)
 -\log(4).
\end{equation}
The product factor in the first braces of~(\ref{7d}) is to be understood
as unity for $p<2$ and in the final brace pair the coefficients of
$\tau^q$
are polynomials in $N$ of degree $ \le q.$
Furthermore, with the $\sqrt{s}$
 factor extracted explicitly as in~(\ref{7d}) these coefficients of
$\tau^q$ vanish if
$q=2k+1$ with $k<p.$ The critical correlation factor in~(\ref{7d}) is
\begin{eqnarray}\label{7f}
 C(N,N,\tau=0)& =& \prod_{n=1}^N \Gamma^2(n)/(\Gamma(n+1/2)\Gamma(n-1/2)), 
\end{eqnarray}
and approaches $ A/N^{1/4}$
as $N \rightarrow \infty$ where \cite{wmtb} $ \log(A) = 3\zeta'(-1)+\log(2)/12$.
We have used~(\ref{7d}), extended or truncated to some order in
$\tau$,
as input to the quadratic recursion formulae to generate all
correlation products $\sqrt{s} C(m,n)$
 as series in $\tau$ within an octant $ m \ge n \ge 0,$
$m+n \le 2N+1.$ While most of our results are numerical, they are
consistent with the assumption that the structure observed on the
diagonal remains true on the octant. That is, if we define $n = \mu
N,$
$0 \le \mu \le 1$ and set $m+n=2N$ (even shell) or $m+n=2N+1$
 (odd shell) then on
these even/odd shells for large $N,$ the correlations are of the form
\begin{equation}\label{7g}
\sqrt{s} C_{\rm e/o}(\mu) = N^{-1/4}\sum_{p=0}^\infty (\log |\tau|
+ \mathcal{L}_N)^p(N\tau/4)^{p^2} A_{\rm e/o}^{(p)}(\mu)
\end{equation}
where the $A_{\rm e/o}^{(p)}(\mu)$
are still Taylor series in $\tau$ as in~(\ref{7d}) but with
coefficients that are now (possibly asymptotic) Laurent series in $1/N$
rather than polynomial in $N.$ The highest power of $N$
multiplying $\tau^q$ remains $N^q.$

 Considerable care must be exercised in the numerical work since
the recursion formulae are unstable. A toy recursion relation of
structure similar to the ones we use in the Ising study illustrates
this nicely. Let

\begin{equation}\label{7h}
C^n_{m+1} = (2(C^n_m)^2 - C_m^{n-1}C_m^{n+1})/C_{m-1}^n
\end{equation}
which is to be applied to all possible $n$ and iterated forward in $m.$
The recursion~(\ref{7h}) has as a solution a constant, say $M,$ but is
susceptible to a steady state growth of errors so that

\begin{equation}\label{7i}
C_m^n \approx M + \epsilon(-1)^n\alpha^m
\end{equation}
is also a possible solution. On substituting~(\ref{7i}) into~(\ref{7h})
one finds the growth constant $\alpha$ must satisfy
$\alpha^2 -6\alpha +1 =0$ so that

\begin{equation}\label{7j}
\alpha = 3+\sqrt{8}, \mbox{ } \log_{10}(\alpha) = 0.765... .
\end{equation}
This is very close to what we find in our numerical work and implies
that if we want some number of digits $D$ that are accurate at the
outer edge of the octant on a shell specified by $n+m=2N+1$ we need to
start with digits $D_0 \approx D + 1.53N.$
In practice we have worked as high
as $N=146$ with $D_0=380$ using the automatic multiple precision facility
of Maple. In section~\ref{sect:sdc} we use these results to 
calculate the ``short-distance'' (including analytic background) terms in the
susceptibility.

\section{Conjectured short-distance structure}\label{sect:qseries}

In this section we state some conjectures for the short distance
behavior of the $\lambda\ne1$ model introduced in
section~\ref{sect:difference}. We arrived at these conjectures
by inspection of series obtained from a combination of the
integrals~(\ref{ci}) and the difference
equations~(\ref{pd1})--(\ref{pd3}).
Our interest in this model is motivated by several considerations.
Firstly, it enables us
to see how the analytic structure of $\chi$ evolves as
successive contributions $\ch{2j}$ or $\ch{2j+1}$ are added.
Secondly, the correlation function $C(M,N;\lambda)$ for small values
of $M$ and $N$ may be required as initial conditions for certain
series generation algorithms. Thirdly, we hope that the presence
of an additional parameter which can be varied will provide some
insight into the Ising model itself. Finally, the deformations of
the elliptic functions that appear in our conjectures may be of
intrinsic mathematical interest.

Following the lead of ref.~\cite{bsw} we make a change of variable from
the modulus, $k$, to the nome $q$. 
Examination of the $j$-particle
contributions to the isotropic susceptibility and two-point functions
reveals that there is much regular structure. 
We arrive at
exact conjectures for $\Ch{j}(0,0)$,
$\Ch{j}(1,0)$, $\Ch{j}(1,1)$, $\Ch{j}(2,0)$ and $\Ch{j}(2,1)$ 
as functions of $j$ and $q$. These provide the first
terms in the short distance expansion of the susceptibility.

\subsection{$q$-series in the Ising susceptibility}
Although more complete results have been obtained for the correlation
functions, we will demonstrate the method by which we derived our conjectures
using the susceptibility as an example.
To make our observations more concrete,
we reproduce tables of series coefficients for
$(1-k^2)^{1/4}\chiso{j}$. As an example of how to interpret the
table, we read $(1-k^2)^{1/4}\chiso{3}=8(k^{9/2}/2^9+4k^6/2^{12}+
16k^{13/2}/2^{13}+4k^7/2^{14}+20k^{15/2}/2^{15}+84k^8/2^{16}+\ldots).$

\begin{table}
\begin{center}
\begin{tabular}{|l|l|l|l|l|l|l|}
\hline
         & 1 & 76 & 1960 & 41888 & 825440 & 15542912 \\
$j=1$    & 4 & 176 & 4256 & 88704 & 1724800 & 32209408 \\
         & 12 & 400 & 9184 &  187264 & 3597440 & 66665984 \\
         & 32 & 896 & 19712 & 394240 & 7490560 & 137826304 \\
\hline
         & 1 & 26 & 556 & 10956 & 206276 & 3772216 \\
$j=2$    & 0 & 0 & 0 & 0 & 0 & 0 \\
         & 4 & 104 & 2224 &  43824 & 825104 & 15088864 \\
         & 0 & 0 & 0 & 0 & 0 & 0 \\
\hline
         & 1 & 16 & 247 & 4140 & 70128 & 1190728 \\
$j=3$    & 0 & 4 & 188 & 4584 & 93456 & 1788648 \\
         & 0 & 20 & 536 &  11164 & 217124 & 4019068 \\
         & 4 & 84 & 1524 & 27884 & 500996 & 8857404 \\
\hline
         & 1 & 34 & 816 & 17032 & 330410 & 6133502 \\
$j=4$    & 0 & 0 & 0 & 0 & 0 & 0 \\
         & 0 & 4 & 184 & 5528 & 137616 & 3080684 \\
         & 0 & 0 & 0 & 0 & 0 & 0 \\
\hline
         & 1 & 48 & 1463 & 36304 & 801661 & 16438116 \\
$j=5$    & 0 & 4 & 228 & 7972 & 221532 & 5382792 \\
         & 0 & 0 & 28 & 1864 & 74112 & 2295212 \\
         & 0 & 4 & 248 & 9468 & 286404 & 7530952 \\
\hline
         & 1 & 70 & 2908 & 93600 & 2582208 & 64243876 \\
$j=6$    & 0 & 0 & 0 & 0 & 0 & 0 \\
         & 0 & 4 & 324 & 15236 & 545744 & 16530604 \\
         & 0 & 0 & 0 & 0 & 0 & 0 \\
\hline
         & 1 & 96 & 5231 & 213136 & 7232113 & 216135776 \\
$j=7$    & 0 & 0 & 4 & 456 & 28952 & 1353328 \\
         & 0 & 0 & 0 & 36 & 4408 & 298448 \\
         & 0 & 4 & 436 & 26588 & 1198004 & 44506752 \\
\hline
\end{tabular}
\caption{Coefficients of the series $(1-k^2)^{1/4}\chiso{j}/2^j$. The numbers
in the tables are the coefficients of $(\sqrt{k}/2)^n$ starting at $n=j^2$
in the upper left corner and reading down and to the right.}
\label{table:summary}%
\end{center}
\end{table}
\begin{table}
\begin{center}
\begin{tabular}{|l|l|l|l|l|l|l|}
\hline
         & 1 & 126 & 8760 & 444740 & 18429842 & 661181352 \\
$j=8$    & 0 & 0 & 0 & 0 & 0 & 0 \\
         & 0 & 0 & 4 & 584 & 46440 & 2666700 \\
         & 0 & 0 & 0 & 0 & 0 & 0 \\
\hline
         & 1 & 160 & 13839 & 858704 & 42821009 & 1823591632 \\
$j=9$    & 0 & 0 & 4 & 708 & 67252 & 4553260 \\
         & 0 & 0 & 0 & 0 & 44 & 8552 \\
         & 0 & 0 & 4 & 728 & 70976 & 4924124 \\
\hline
         & 1 & 198 & 20888 & 1560492 & 92610504 & 4644898080 \\
$j=10$   & 0 & 0 & 0 & 0 & 0 & 0 \\
         & 0 & 0 & 4 & 868 & 99812 & 8087916 \\
         & 0 & 0 & 0 & 0 & 0 & 0 \\
\hline
         & 1 & 240 & 30359 & 2692864 & 188045229 & 11006289872 \\
$j=11$   & 0 & 0   & 0     & 4       & 1064      & 148424 \\
         & 0 & 0   & 0     & 0       & 0         & 52 \\
         & 0 & 0   & 4     & 1044    & 143020    & 13686020 \\
\hline
         & 1 & 286 & 42756 & 4447860 & 361695338 & 24490780096 \\
$j=12$   & 0 & 0   & 0     & 0       & 0         & 0 \\
         & 0 & 0   & 0     & 4       & 1256      & 205352 \\
         & 0 & 0   & 0     & 0       & 0         & 0 \\
\hline
         & 1 & 336 & 58631 & 7076256 & 663817077 & 51575531568 \\
$j=13$   & 0 & 0   & 0     & 4       & 1444      & 270020 \\
         & 0 & 0   & 0     & 0       & 0         & 0 \\
         & 0 & 0   & 0     & 4       & 1464      & 277424 \\
\hline
         & 1 & 390 & 78584 & 10898664 & 1169440708 & 103475590040 \\
$j=14$   & 0 & 0   & 0     & 0        & 0          & 0 \\
         & 0 & 0   & 0     & 4        & 1668       & 358612 \\
         & 0 & 0   & 0     & 0        & 0          & 0 \\
\hline
\end{tabular}
\caption{Table continued.}
\label{table:summaryb}%
\end{center}
\end{table}

Inspecting Table~\ref{table:summary},
we make the conjecture that the series can be sensibly decomposed as
\begin{multline}
2^j (\sqrt{k}/2)^{j^2}[(1+c_{0,1}k^2+c_{0,2}k^4+\ldots) \\
+(\sqrt{k}/2)^j(4+k+c_{1,2}k^2+c_{1,3}k^3+\ldots) \\
+(\sqrt{k}/2)^{2j}(c_{2,0}+c_{2,1}k+c_{2,2}k^2+\ldots)+\ldots]
\label{decomp}
\end{multline}

In fact, we observe that as $j$ tends to larger and larger values,
the first row of coefficients
tends towards the expansion in $k$ of
$2^jq^{j^2/4}/\theta_3(0,q)$.\footnote{This may be checked by reverting the
product form given in~(\ref{2b}) for $k=k(q)$ and using the identity
$\theta_3(0,q)=\sqrt{2{\rm K}/\pi}$ to obtain the expansion of $\theta_3(0,q)$
in $k$.} The first correction comes in at order $q^{j(j+1)/4}$.
Hence we make the change of variable from $k$ to $q$ in $\chiso{j}$ and divide
the result by $2^j q^{j^2/4}/\theta_3(0,q)$. For all $j$ we obtain a series
of the form $1+4q^{j/4}+cq^{(j+1)/4}+\ldots$
The sequence of terms starting at order $q^{j/4}$
appears again to be fitted by a recognizable function of $q$,
at least until contributions appear at order $q^{2j/4}$
or $q^{3j/4}$.
Subtracting this assumed product form yields a series whose
first correction enters at order $q^{2j/4}$. The coefficients of the
terms of orders lying between $q^{2j/4}$ and $q^{3j/4}$ are not independent
of $j$ as was the case previously, but vary linearly with $j$. The
constant part has a product form, and the $j$-dependent part may as well
be we do not have sufficiently many terms to make a firm conjecture.
Likewise the terms between $q^{3j/4}$ and $q^{4j/4}$ depend quadratically
on $j$, and the correction at $q^{4j/4}$ appears to vary as the fourth
power of $j$.

From the long series we have produced,
we have been able to conjecture that
\begin{equation}\label{split}
\begin{split}
&2^{-j}(1-k^2)^{1/4}\chiso{j}=
\frac{q^{j^2/4}}{\theta_3(0,q)} \left[ 1+4q^{j/4}\prod_{n=1}^\infty
\frac{1+q^{n-1/2}}{1+q^n}\right.\\
&\qquad+4q^{2j/4}\left(\prod_{n=1}^\infty(1-q^{2n-1})^4
(1+q^{n-1/2})^4+(j+1)4q^{1/4}/\theta^2_2(0,q^{1/2})\right)\\
&\left.\qquad+4q^{3j/4}\left(f^{(0)}_3(q)+(j+1)(j+2)f^{(2)}_3(q)\right)+
O\left[q^{4j/4}\right] \right]
\end{split}
\end{equation}
where
\begin{align}
f^{(0)}_3&=1+6q^{1/2}+26q+17q^{3/2}-81q^2-55q^{5/2}+285q^3\ldots\\
f^{(2)}_3&=1+q^{1/2}-5q-4q^{3/2}+15q^2+10q^{5/2}-39q^3+\ldots
\end{align}
These are consistent with the expressions for the correlation functions
in the following section. That is to say, summing the correlation
functions given in the following subsection gives terms that agree,
as far as they should, with the above expression. Similarly, 
summing~(\ref{split}) over $j$ gives a series that agrees to the appropriate
(low) order with the known expansion for $\chi_{iso.}$

\subsection{$q$-series in the two-point functions}

In this subsection
we write down some conjectured results for the short-distance
correlation functions which we have derived
empirically. For these cases, unlike the susceptibility, the general
correction term is apparent from the series and we are able to formulate
exact conjectures.
Let us define operators, $\Phi_0$ and $\Phi_1$,
which convert power series in $z$ to power series in $q$ according to
the rules
\begin{align}
\Phi_0 \cdot\sum_{n=0}^\infty c_n z^n &= \sum_{n=0}^\infty c_n q^{n^2/4}, \\
\Phi_1 \cdot\sum_{n=0}^\infty c_n z^n &= \sum_{n=0}^\infty c_n q^{n(n+1)/4}.
\end{align}
Then we conjecture that
\begin{align}
&2^{-j}(1-k^2)^{1/4}\Ch{j}(0,0)=
\frac{1}{\theta_3(0,q)}\Phi_0 \frac{z^j(1-z^2)}{(1+z^2)^{j+1}},\\
&2^{-j}(1-k^2)^{1/4}\Ch{j}(1,0)=
\frac{1}{\theta_3(0,q)}\left(\frac{2q^{1/8}\theta_3(0,q^{1/2})}
{\theta_2(0,q^{1/2})}\right)^{1/2} \Phi_1 \frac{z^j(1-z)}{(1+z^2)^{j+1}},\\
&2^{-j}(1-k^2)^{1/4}\Ch{j}(1,1)=\frac{2(j+1)}{\theta_2(0,q)\theta_3^2(0,q)}
\Phi_0 \frac{z^{j+1}(1-z^2)}{(1+z^2)^{j+2}},\\
&2^{-j}(1-k^2)^{1/4}\Ch{j}(2,0)=\frac{1}{q^{1/4}\theta_3(0,q)}
\left(\frac{2q^{1/8}\theta_3(0,q^{1/2})}{\theta_2(0,q^{1/2})}\right)^{4/2}
\Phi_0 \frac{z^{j+1}}{(1+z^2)^{j+1}(1-z^2)} \nonumber\\
&\qquad-\frac{16}{\theta_3(0,q)\theta_2^4(0,q^{1/2})}q^{1/4}\frac{d}{dq^{1/4}}
\left[\Phi_0\frac{z^{j+2}}{(1+z^2)^{j+2}(1-z^2)}\right]
-4\left[\frac{1}{\theta_3(0,q)}\right.\nonumber\\
&\qquad\left.+\frac{\theta_2(0,q)}{2\theta_3^2(0,q)\theta_2(0,q^4)}
-\frac{8}{\theta_3(0,q)\theta_2^4(0,q^{1/2})\theta_2(0,q^4)}
q^{1/4}\frac{d}{dq^{1/4}}\left[\Phi_0\frac{z^2}{(1+z^2)^2(1-z^2)}\right]
\right]\nonumber\\
&\qquad\times\Phi_0\frac{z^{j+2}}{(1+z^2)^{j+2}(1-z^2)},\\
\text{and}\nonumber\\
&2^{-j}(1-k^2)^{1/4}\Ch{j}(2,1)=\frac{1}{\theta_3(0,q)}
\left(8q^{5/16}\frac{\theta_2(0,q^{1/4})}{\theta_2^5(0,q^{1/2})}
\frac{d}{dq^{1/4}}\left[\Phi_1\frac{z^{j+1}(1+z)}{(1+z^2)^{j+1}(1-z^2)}\right]
\right.\nonumber\\
&\qquad+\left(\Phi_1\frac{z(1+z)}{1-z^4}\right)^{-1}
\Phi_1\frac{z^{j+1}(1+z)}{(1+z^2)^{j+1}(1-z^2)}\nonumber\\
&\qquad\left.\times\left[\theta_4(0,q)-
8q^{5/16}\frac{\theta_2(0,q^{1/4})}{\theta_2^5(0,q^{1/2})}
\frac{d}{dq^{1/4}}\left[\Phi_1\frac{z(1+z)}{1-z^4}\right]\right]
\right)
\end{align}

As noted above, these are necessary, but not sufficient, for the generation 
of the $j$-particle contributions to the correlation functions, being
some of the initial conditions for the recurrences.

\section{Scaling form of the susceptibility.}\label{sect:numerical}

The main result in this section is a conjecture completely
specifying the analytic structure of the susceptibility $\chi$ of the
isotropic Ising model in the vicinity of the critical point both in
the neighborhood of the ferromagnetic point $s=1$ and the
anti-ferromagnetic point $s=-1$. The conjecture, contained in
eqs.~(\ref{6f}--\ref{9d}) is based on what we believe is overwhelming numerical
evidence that is obtained by disentangling the ``short- distance" and
``scaling" parts of $\chi$ in a manner described below.

In section~\ref{sect:sdc} we give the assumptions and numerical procedures 
we use to derive the ``short-distance" part of $\chi$ in~(\ref{6f}) 
with the coefficients listed in the Appendix. Then, in section~\ref{sect:anal}
we describe the ``short-distance" subtraction and analysis on which 
the behavior of the scaling-amplitude function $F_{\pm}$ shown
in~(\ref{9d}) is based.
Finally, in section~\ref{sect:ca} we outline our fitting programs to determine
the coefficients in the functions $F_{\pm}$. Since there are no confluent
singularities in $F_{\pm}$ whose amplitudes need to be found, the fitting
procedure is very well-conditioned and the coefficients in the Appendix 
are as determined to an accuracy of up to 20 digits.

\subsection{``Short-distance'' term.}\label{sect:sdc}

That the ``short-distance'' contribution to $\chi$
can be obtained 
from numerical values of $C(m,n)$ for small $|m|$ and $|n|$ relies on
certain assumptions about the behavior of the expansion coefficients
in eq.~(\ref{7g}). In particular, we assume that (\ref{7g}) remains valid
up to $N$ of the order $1/\tau$ where it can, in principle, be matched
term by term to a large distance expansion that properly describes
the roughly exponential $\exp(-N \tau)$ decay of correlations as
$ N \rightarrow \infty.$ Explicit matching formed the basis of the previous
calculations of terms in the ``short-distance" $\chi$ (cf. \cite{kapb})
but this becomes extremely cumbersome at higher order. Our ability to
go to high order here rests on the fact that we dispense with such
explicit matching and rely instead on power counting to uniquely
identify those terms that contribute to the ``scaling" and the 
``short-distance" parts of $\chi$ separately. We believe this is tantamount to
the scaling argument that in the critical region there is a single
length scale proportional to $1/\tau^\nu$ with $\nu=1$
 and thus, in a way to
be made more precise below, we can deduce the power law of the large
distance contribution of any set of terms varying as $N^p$ by simply
replacing $N^p$ with $1/\tau^p.$ Terms whose variation is as a
fractional power of $\tau$ (with possibly logarithmic multipliers) are
discarded as assumed contributions to the ``scaling" part of $\chi.$
Terms whose variation is predicted to be an integer power of $\tau$
(with possibly logarithmic multipliers) are assumed to be part of the
``short-distance" $\chi$ and are treated more carefully.

To make the argument and assumptions more explicit we begin
with some definitions that will be useful also for the subsequent
analysis. Let the two dimensional sum (\ref{1.8}) defining $\chi$ be reduced
to a one dimensional sum by
combining the contributions from all sites on the even and odd
squares $|m|+|n| = 2N$, $2N+1$
 and then further combining these into sum
and difference combinations which are necessary for separating the
ferromagnetic and anti-ferromagnetic contributions. That is we write
\begin{equation}\label{7k}
\sqrt{s}C_{N_\pm} = \sum_k C_{N_\pm}^{(k)}\tau^k =
 \sqrt{s}\sum (C_{\rm e}(\mu) \pm C_{\rm o}(\mu))
\end{equation}
where the first sum simply defines the coefficients $ C_{N_\pm}^{(k)}$ while
the second is the actual lattice sum over the octant correlations defined
by~(\ref{7g}) and extended by symmetry over the full square. The
coefficients $A_{N_\pm}^{(l,k)}$
in the expansion

\begin{equation}\label{7l}
C_{N_\pm}^{(k)} = \sum_{l=0}^L A_{N_\pm}^{(l,k)} (\log{|\tau|} + \mathcal{L}_N)^l
\end{equation}
have the large $N$ asymptotic expansions
\begin{equation}\label{7m}
A_{N,{\rm f}}^{(l,k)}= \sum_{p=0} A_{\rm f}^{(l,k,p)}N^{3/4+k-p},\mbox{ }
A_{N,{\rm af}}^{(l,k)}= \sum_{p=0} A_{\rm af}^{(l,k,p)}N^{-1/4+k-p},
\end{equation}
as assumed in~(\ref{7g}) based on numerical evidence. The ferromagnetic
and anti-ferromagnetic cases have been treated separately in~(\ref{7m})
to emphasize that shell subtraction in~(\ref{7k}) reduces the leading
power of $N$ by unity. The upper limit $L$ on the logarithmic powers 
in~(\ref{7l}) depends on $k$ as discussed in connection with
eq.~(\ref{7d}) but its
precise value is not needed in the following discussion. We also
introduce the partial sums
\begin{eqnarray}\label{7n}
S_{N_\pm}^{(k)} = \sum^N_{n=0} C_{n_\pm}^{(k)} =\sum_{l=0}^L
 R_{N_\pm}^{(l,k)}(\log{|\tau|)^l} 
&=&\sum_{l=0}^L (\beta_{\pm}^{(l,k)} +
 \delta R_{N_\pm}^{(l,k)})(\log{|\tau|)^l}
\end{eqnarray}
where the $ R_{N_\pm}^{(l,k)}$
in the second equality in~(\ref{7n}) are numerical coefficients
that are directly generated by the quadratic recursion relations and
the subsequent lattice summations. In the final equality in~(\ref{7n})
these expansion coefficients have been formally split into two terms. 
This formal separation is to be understood in an asymptotic sense with
$\beta_{\pm}^{(l,k)}$ a constant that is the $N$ independent part of
$R_{N_\pm}^{(l,k)}$ as $N \rightarrow \infty$
 and which we can understand as an ``integration constant".
 The remainder $\delta R_{N_\pm}^{(l,k)}$, we will show in the case
 that $l=L$, has an
asymptotic expansion like the $ A_{N_\pm}^{(l,k)}$ in~(\ref{7l})
except for an extra factor of $N$ from the summation. For $l \ne L$
the $\delta R_{N_\pm}^{(l,k)}$ can be expressed as sums of such
expansions with multiplying logarithmic factors $\mathcal{L}_N.$

We now come to our key assumption. The partial sums in (\ref{7n}) in
most cases diverge as $N \rightarrow \infty$
because of the presence of
large fractional powers of $N$ in the $\delta R_{N_\pm}^{(l,k)}$.
However, we assume that if the formal matching of the short and large
distance expansions for $C(m,n)$ had been carried out, these partial
sums would in fact converge and furthermore the contribution of each
power of $N$
term could be estimated by the replacement $N \rightarrow 1/\tau.$
Since all the powers of $N$ in $\delta R_{N_\pm}^{(l,k)}$ are
fractional, the result is that these terms contribute only to the
``scaling" part of $\chi.$ Thus the ``short-distance" part of 
$\chi$ comes
entirely from the ``integration constant'' term $\beta_{\pm}^{(l,k)}$
in (\ref{7n}), and on explicitly reinserting the factors of $\sqrt{s}$
and $\tau^k$ we get
\begin{equation}\label{7o}
B_{\rm f/af} = \sum_k\sum_{l=0}^L \beta_{\pm}^{(l,k)}\tau^k
 (\log{|\tau|)^l}/\sqrt{s}
= \sum_{p=0}^\infty \sum_q b_{\rm f/af}^{(p,q)} \tau^q (\log{|\tau|)^p}
\end{equation}
at the ferromagnetic and anti-ferromagnetic points. The last equality
in~(\ref{7o}) is eq.~(\ref{6f}) and follows simply by expanding $\sqrt{s}$
in a series in $\tau.$ The rest of this section comprises
a discussion of the numerical scheme we use to isolate the
``integration constants'' efficiently.

A technical problem arises in that one must somehow isolate the
constant $\beta_{\pm}^{(l,k)}$
from an $N$-dependent sequence $R_{N_\pm}^{(l,k)}$ and this can be a
very unstable procedure if the $R_{N_\pm}^{(l,k)}$ are combinations involving
logarithms. Fortunately there are no logarithms in $R_{N_\pm}^{(L,k)}$
and one can
iteratively remove the logarithmic factors in the remaining $R_{N_\pm}^{(l,k)}$
by working in sequence from $l=L, L-1, ...$ to $l=0$. To understand 
precisely how to do the subtractions we first relate the unknown 
$N-$dependent structure of the $R_{N_\pm}^{(l,k)}$
to the known, or rather assumed,
simple structure of the $A_{N_\pm}^{(l,k)}.$ From the definition
of $S_{N_\pm}^{(k)}$ as a
partial sum it follows that $S_{N_\pm}^{(k)} - S_{N-1_\pm}^{(k)}
= C_{N_\pm}^{(k)}$ and by comparing
coefficients in eqs.~(\ref{7n}) and~(\ref{7l}) we get
\begin{equation}\label{7p}
R_{N_\pm}^{(l,k)} - R_{N-1_\pm}^{(l,k)} =
 \sum_{m=l}^L \binom{m}{l} A_{N_\pm}^{(m,k)}
(\mathcal{L}_N)^{m-l}.
\end{equation}
In the case that $l=L$ there are no logarithmic factors on the right
hand side of~(\ref{7p})
and the equation is just a discrete first order
differential equation in $N$ whose solution is an integration constant
additive to an asymptotic series of the same form as in~(\ref{7m}) except
for an extra power of $N.$ The easiest way to determine the
``integration constant'' $\beta_{\pm}^{(L,k)}$ is to fit it together with
unknown coefficients defining the asymptotic series part to a sequence of
$R_{N_\pm}^{(L,k)}$. This is a stable numerical procedure,
although if $k$ is large,
$\beta_{\pm}^{(L,k)}$ is sub-dominant to many much larger terms and it is
essential to calculate in high precision. For example, we might start the
recursive calculation of $R_{N_\pm}^{(L,k)}$
as described in the previous section
with $D_0=380$ digits, end with $D=155$ digits at $N=146,$ and continue with
this number of digits in a $71 \times 71$ matrix inversion to obtain the
$\beta_{\pm}^{(L,k)}$ of interest while discarding the remaining 70
coefficients of the asymptotic series! The results are slightly more accurate
than the final answers that are given in the Appendix.

For $l \ne L$ the presence of logarithmic factors on the right-hand
side of~(\ref{7p}) makes fitting to $R_{N_\pm}^{(l,k)}$
impractical. Instead we define
subtracted functions

\begin{equation}\label{7q}
F_{N_\pm}^{(l,k)} = R_{N_\pm}^{(l,k)} +
 \sum_{m=l+1}^L \binom{m}{l}(R_{N_\pm}^{(m,k)}-\beta_{\pm}^{(m,k)})
(-\mathcal{L}_N)^{m-l}
\end{equation}
which we use as replacements for $R_{N_\pm}^{(l,k)}$ in all fitting procedures. 
That the fitting functions $F_{N_\pm}^{(l,k)}$
 will give the same $\beta_{\pm}^{(l,k)}$ is obvious
since the added terms in~(\ref{7q}) all contain logarithmic factors and
are thus not $N$-independent. That the $F_{N_\pm}^{(l,k)}$
 are logarithm free follows from

\begin{equation}\label{7r}
F_{N_\pm}^{(l,k)} -F_{N-1_\pm}^{(l,k)}= A_{N_\pm}^{(l,k)} +
 \sum_{m=l+1}^L \binom{m}{l}(F_{N-1_\pm}^{(m,k)}-\beta_{\pm}^{(m,k)})
(\mathcal{L}_{N-1} - \mathcal{L}_N)^{m-l}
\end{equation}
and an argument by induction starting from $l=L$. Eq.~(\ref{7r}) can be
verified using the definition~(\ref{7q}) and the subtraction eq.~(\ref{7p}).

To summarize, we obtain
$\beta_{\pm}^{(l,k)}$ numerically by using~(\ref{7q}) iteratively,
starting with $l=L-1,$ to generate $F_{N_\pm}^{(l,k)} $ using
procedures similar to that described above for the initial
$F_{N_\pm}^{(L,k)}$ which are just the $R_{N_\pm}^{(L,k)}.$ The
results are given in the Appendix.

Our arguments above are at best plausible and to provide a
rigorous proof justifying our numerical procedure we believe one
would have to do three things. First, one would have to show the
$C(m,n)$ expansion has the form (\ref{7g}) we deduced on numerical grounds.
Second, one would have to show there exists a corresponding
asymptotic expansion valid at large $N.$ Third, one must show both
expansions have a sufficiently large domain of validity that the
matching we assumed could in fact be carried out to arbitrarily high
order. On the other hand, we have numerical evidence for the
validity of our procedure as described in the next section. The
``short-distance" terms we have calculated, when used in a series
subtraction process, leave a residual in high order series
coefficients that is consistent with the complete elimination of all
``short-distance'' terms, both singular and analytic, to the 
O$(\tau^{14})$ we have worked. While the main intent of section
\ref{sect:anal} is actually quite different, it is highly unlikely that the
cancellations necessary to yield the expansion of the
scaling-amplitude function $ F_{\pm}$ of eq.~(\ref{9d}) in pure integer powers
of $\tau$ would have occurred had there been an error, numerical or
otherwise, in the results of this section.

We conclude this section with a toy model example to illustrate
the possible convergence properties of the short-distance expansion.
The main impediment to extending the calculation described above is
that we do not have simple analytical expressions for the $\tau$
expansion of the $C(M,N)$ in general. An exception is on the diagonal
where we know, cf. eq.~(\ref{7d}), that the coefficient of 
$\tau^{p^2} (\log|\tau|)^p$ is
\begin{equation}\label{t9}
C_N^{(p)} = 4^{p-p^2} C(N,N,\tau=0) N^p
  \prod_{k=1}^{p-1} (1-N^2/k^2)^{p-k}
\end{equation}
and a simple expression for the asymptotic expansion of $C(N,N,\tau=0)$
at large $N$ can be found in Au-Yang and Perk \cite{ap84}.
We can define the diagonal partial sums 
$S_N^{(p)} = \sum_{n=0}^N C_n^{(p)}$
 and, as in the analysis described above, ask
for the short-distance coefficient $b_{\text{diag}}^{(p)}$ which is the $N$
independent part of $S_N^{(p)}$ in the limit $N \rightarrow \infty.$
This term $b_{\text{diag}}^{(p)}$ could be viewed as a partial
contribution to the
short-distance terms of interest but since we do not know what
cancellations will occur when we include all $C(M,N),$ we prefer to
consider it only as a toy result that is suggestive for the
convergence of the short-distance terms with order $p.$

The term $b_{\text{diag}}^{(p)}$ is also equal to the $\epsilon$
independent part of
$\sum_{n=0}^{\infty} C_n^{(p)} e^{-\epsilon n}$ in the limit $\epsilon
\rightarrow 0.$ This latter expression is more convenient since we
can add to it any term such as $1/(1-e^{-\epsilon})^{p^2+3/4-k}$ for
integer $k$ without contributing to any $\epsilon$ independent term. If
we expand such terms in series in $e^{-\epsilon},$ we obtain as an
equivalent sum
\begin{equation}\label{t10}
\sum_{n=0}^{\infty} \left[ C_n^{(p)}
  - \sum_{k=0}^K g_k \Gamma(n+p^2+3/4-k)/n! \right] e^{-\epsilon n}.
\end{equation}

Now choose the $g_k$ in (6.10) such that the divergent terms in the
asymptotic $n$ expansion of $C_n^{(p)}$ cancel the divergent terms in the
Gamma functions. Then the $n$ sum becomes convergent even with $\epsilon = 0.$
With the cancellation extended to include also some slowly
decaying terms in $n,$ thus requiring $K > p^2,$ one obtains the explicit
formula
\begin{equation}\label{t11}
b_{\text{diag}}^{(p)} = \sum_{n=0}^{\infty} \left[ C_n^{(p)}
 - \sum_{k=0}^K g_k \Gamma(n+p^2+3/4-k)/n! \right]
\end{equation}
which is very convenient for numerical work. We find the
$b_{\text{diag}}^{(p)}$ calculated from (\ref{t11}) for $p=1,2,\ldots,20$ are

\begin{align}\label{t12}
  -1.30&\times 10^{-1},  & 4.29&\times 10^{-4},  & 1.58&\times 10^{-5},
   & -9.12&\times 10^{-9}, \nonumber \\
  -6.31&\times 10^{-11},  & 8.31&\times 10^{-14}, & 2.38&\times 10^{-15},
   & -3.73&\times 10^{-17}, \nonumber \\
  -2.08&\times 10^{-17}, & 1.18&\times 10^{-17}, & 3.61&\times 10^{-16},
   & -1.78&\times 10^{-14}, \\
  -6.67&\times 10^{-11}, & 5.80&\times 10^{-7},  & 5.15&\times 10^{-1},
   & -1.44&\times 10^{+6}, \nonumber \\
  -5.34&\times 10^{+14}, & 8.13&\times 10^{+23}, & 2.06&\times 10^{+35},
   & -2.70&\times 10^{+47}. \nonumber
\end{align}

The sequence in (\ref{t12}) clearly shows the asymptotic nature of the toy
expansion. Furthermore the magnitude of the terms is in semi-quantitative 
agreement with $\Gamma(p^2/2)/a^{p^2/2}$ from eq.~(\ref{eqcut}).
Thus while we cannot conclude that this will be how the short-distance
susceptibility terms $B_{\rm f/af}$ in (\ref{6f}) will behave, it is at least
encouraging to note that there is no evidence for any behavior more
singular than that predicted by our natural boundary analysis.

\subsection { Series ``proof'' of $F_{\pm}$ behavior}\label{sect:anal}

Any numerical analysis program to deduce functions such as the
scaling-amplitude functions $F_{\pm}$ from their series expansions is
in essence a fitting
routine and always presupposes a knowledge of the analytic structure
of the result. Although we have very little exact knowledge of $F_{\pm}$
we do know that the multiplier of $|\tau|^{-7/4}$ in $\cu{2}$ (cf.
eq.~(\ref{n2})) contains $\log{|\tau|}$ terms and there is numerical evidence
\cite{nica} that this is true of the $\cu{n}$ for $n>2$ also. On the other
hand, Gartenhaus and McCullough \cite{GM} found that the series for $\chi$ were
consistent with the absence of logarithmic corrections in $F_+$ through
order $\tau^3$ and this was subsequently confirmed for $F_-$ as well
\cite{nicb}. Scaling arguments on the question of logarithmic
corrections are necessarily inconclusive because of the lack of
information on amplitudes, some of which may vanish.
We know of no scaling argument that either definitely
requires the presence of logarithmic terms or can definitely exclude
them.

The natural boundary arguments given in section \ref{sect:circle} preclude
the possibility that $\chi$ has a convergent rather than asymptotic
expansion about $\tau = 0$, but we know of no analytical argument that
shows whether this applies to both the ``scaling'' and ``short-distance''
terms or just to one or the other.\footnote{A simple ratio analysis of
the available terms in the expansion of $\sqrt{s}F_{\pm}$ in the
variable $\tau^2$ (see appendix) is indicative of a convergent series
with radius of convergence $\approx1.2$. This implies a conjugate
pair of singularities at $\tau\approx1.1i$. If however the series is
asymptotic, this observation will fail to hold.}
Neither could we distinguish the convergent from the asymptotic expansions
from the data we have. Although there are singularities in $\chi$ dense
on the line $-i \le \tau \le i$, the singularities close to $\tau=0$
are extremely weak and we have not detected any singularity closer to
the ferromagnetic $\tau=0$ point than that arising out of 
$\ch{6}$ at $\Im(\tau) = \sqrt{7}/4 \approx 0.66.$
Thus for all practical (numerical) purposes we can ignore the possible
asymptotic nature of the ``scaling'' terms.

The possibility that the $F_{\pm}$ might be expanded in a series in
{\bf integer} powers of $\tau$ can be confirmed numerically without
evaluating any of the coefficients in the expansion. The
trick is to generate the series for the susceptibilities divided by
the factor $(1-k^2)^{1/4}$. Any term in $F_+$ or $F_-$ that is
{\bf not} a pure integer power law $\tau^p$, $p>1$, will necessarily
contribute to the high order coefficients of the series. Of course,
the leading two terms $1+\tau/2$ in $F_{\pm}$, which occur as poles
proportional to $1/\tau^2$ and $1/\tau$ in the scaled $\chi_{\pm}$, will also
contribute, but the amplitude of these terms is known and so their
contribution can be subtracted. Similarly, given the accurate
``short-distance'' amplitudes in the Appendix, the contribution of all
``short-distance'' terms can be similarly eliminated through order
$\tau^{14}.$ We find that the remaining high order series coefficients are
plausibly consistent with the ``short-distance'' ${\rm O}(\tau^{15})$ that has
not been subtracted and thus that
there is {\bf no numerical evidence} for any powers of $\tau$
other than pure integers
in the scaling-amplitude functions $F_{\pm}$. The rest of this
section gives details of the analysis that is the basis for this
conclusion while the following section reports on our procedures for
estimating the coefficients in the $\tau$ expansions of $F_{\pm}$.

As outlined above, our search for possible terms other than those
with pure integer
powers of $\tau$ in the $F_{\pm}$ involves the observation of the high 
order series terms in
the scaled and pole subtracted susceptibility functions

\begin{equation}\label{8a}
\Delta\chi_+ = \beta^{-1} \chi_+/(1-s^4)^{1/4}
 -(2K_c \sqrt{2})^{7/4} C_0^+ 2^{-1/2} /(1-s)^2,
\end{equation}

\begin{equation*}
\Delta\chi_- = \beta^{-1} \chi_-/(1-s^{-4})^{1/4}
 -(2K_c \sqrt{2})^{7/4} C_0^- 2\sqrt{2}s^2 /(s^2 - 1)^2
 = \sum_{m=1}^\infty K^-_{2m} s^{-2m}
\end{equation*}
applicable for $T>T_c$ and $T<T_c$ respectively. Since the procedures for
$ T>T_c$ and $T<T_c$ are not different in principle, we will restrict the
discussion below to the $T<T_c$ case only.

To implement the ``short-distance'' subtraction we can restrict
ourselves to determining $b_n$, the contribution to the coefficient of
$s^{-n}$ arising from the $s^{-1}=1$ singularity in $(1-s^{-4})^{-1/4}
(\log(-\tau))^p/\sqrt{s}.$\footnote{The denominator $\sqrt{s}$ factor here and
in the following is included for the convenience of allowing us to
use the ``short-distance'' amplitudes exactly as tabulated in the Appendix.
It also simplifies the final formula (\ref{715}).} The contribution from
$\tau^q (1-s^{-4})^{-1/4}(\log(-\tau))^p/\sqrt{s}$ follows trivially from
$b_n$ by $q$ repetitions of the derivative operation
$D_{\tau} b_n = (b_{n+1} - b_{n-1})/2$ because of the simple form 
$\tau = (s^{-1} -s)/2.$
Note also
that the contribution from the $1/s=-1$ singularity is identical
except for an overall sign $(-1)^n$ and thus simply requires that in
the end we set all odd power amplitudes to zero and double the even
ones. To determine $b_n$ is a standard exercise in complex variable
contour integration; we deform the contour in an $s^{-1}$ integral with
integrand $s^{n+1}/(2\pi i) (1-s^{-4})^{-1/4}(\log(-\tau))^p/\sqrt{s}$ to
surround the branch-cut $1 \le s^{-1} < \infty$ and in a final step
set $s^{-1} = \exp{x}.$ The result is
\begin{equation}\label{715}
b_n = \pi^{-1} \int_0^\infty dx \exp(-nx) (2\sinh{2x})^{-1/4}
  \Im \{\exp(i\pi/4) (\log{\sinh{x}} -i \pi)^p\}
\end{equation}
and although the integral (\ref{715}) cannot in general be done in closed
form, expansions valid asymptotically for large $n$ are easy to
generate with computer algebra packages such as Maple. Because our
series are particularly long these asymptotic expansions are
essentially exact and are also very convenient for the subsequent
$D_\tau$ differentiations.

It is also necessary to subtract and/or smooth out the
contributions from complex singularities on the circle $|s^{-1}|=1.$
The only significant ones for the present calculation are those at
$s^{-1}=\pm i$ and $\pm\exp{(\pm i \pi/3)}.$ The subtractions of the leading
contributions from the latter singularities arising from $\hat{\chi}^{(4)}$ can
be obtained directly from eq.~(28) in \cite{nicb}. The result is

\begin{align}\label{716}
  K^-_{2m} &\rightarrow  K^-_{2m}+ 1536/(5005\pi) (4 \sqrt{2}/(2m)!)
      \left[3^{1/4} \sin(2m\pi/3 +\pi/4) \{(-13/2)_{2m}  \right.\nonumber\\
  & + 13/4(-15/2)_{2m} +1265/1632(-17/2)_{2m} -5365/384(-19/2)_{2m}\}
     \nonumber \\
  & -3^{-1/4} \cos(2m\pi/3 +\pi/4) \{5/2(-15/2)_{2m} +75/8(-17/2)_{2m}
     \nonumber\\
  & \left. +53321/20672(-19/2)_{2m}\} \right]
\end{align}
where $(z)_n = \Gamma(z + n)/\Gamma(z)$ is a Pochhammer symbol.

We find that the further smoothing to give the remainder

\begin{equation}\label{717}
R^-_n = n^{1/2} D_b^4 n^{-1/2}(n^4 D_a^4)^3 n^3 K^-_n
\end{equation}
is more than adequate. Here as in \cite{nicb} $D_a g_n = (g_{n-1}
+g_{n+1})/2$ suppresses the contributions from $s^{-1}=\pm i$
while $D_b g_n = (g_{n-2} +g_n +g_{n+2})/3$ reduces what has not been
subtracted by (\ref{716}). Note that the smoothing in eq.~(\ref{717}) scales
up the amplitude contributions from the singularities at $s^{-1}=\pm 1$
by a factor $n^{15}.$

We find that as we include higher and higher orders of the
``short-distance'' terms in the subtraction process the remainder
amplitudes (\ref{717}) decrease in a smooth fashion. The amplitude $R^-_n$
we obtain after having subtracted all terms through ${\rm O}(\tau^{14})$ listed
in the Appendix is, at $n=600$, $\approx -4.1\times 10^7$ compared to
$\approx 1.5 \times 10^{43}$ one gets without pole subtraction in
(\ref{8a}). The residual is very reasonably the amplitude we would
expect from the ${\rm O}(\tau^{15})$ terms and this has been confirmed by
extending the ``short-distance'' amplitude sequence in the Appendix by two
terms using a crude Pad\'e analysis. The result of this additional
subtraction is to reduce $R^-_n$ at $n=600$ to $\approx (-1 \text{ to } 1) 
\times 10^6.$
Analysis of the high temperature series leads to a similar conclusion
and also that the anti-ferromagnetic ``short-distance'' terms in the
Appendix represent the susceptibility at this point completely with
nothing left out.

Although our analysis does not ``prove'' the absence of 
powers of $\tau$ other than pure integers, we can put
very stringent bounds on the amplitudes of any possible singular
terms. To make this comparison concrete, we suppose either of $F_{\pm}$
contains the singular term $A_p \tau^p \log|\tau|$ and for simplicity
assume $p$ integer. The contribution of this term to the amplitude of
the coefficient of $s^n$ for $T>T_c$ or $1/s^n$ for $T<T_c$ in the high/low
temperature series expansion of (\ref{8a}) will be about
$A_p\Gamma(p-1)/n^{p}$ relative to the pole contribution and this is to be
compared to the observed amplitude we obtain after subtracting or smoothing
away the known singularities as best we can. Our current results are the
amplitude bounds

\begin{eqnarray}\label{8b}
|A_p| & < & 10^{-35} 300^{p}/\Gamma(p-1), \mbox{ } T>T_c,\\
|A_p| & < & 10^{-37} 600^{p}/\Gamma(p-1), \mbox{ } T<T_c,
\end{eqnarray}
and these bounds essentially exclude any singularity with reasonable
amplitude, scaling as $\tau^p,$ for all $p$ less than about 15.

\subsection{ $F_{\pm}$ coefficient analysis}\label{sect:ca}

The task of determining $F_{\pm}$ numerically is enormously
simplified by the {\em a priori} knowledge---strictly speaking a
conjecture based on the numerical work of the last section---that
$F_{\pm}$ has an expansion in integral powers of $\tau$
near $\tau=0.$ The absence of any
confluent terms means that many different analyses can be used
efficiently and we report here on two independent calculations that
give essentially identical results, thus again confirming the
above conjecture. We have not seriously attempted to optimize
our analysis to give the most accurate numerical values possible and
thus if it ever becomes necessary one could almost certainly improve
on our coefficients as given in the Appendix.

One particular feature of the $F_{\pm}$ expansion is worth noting
here. When the functions are scaled by $\sqrt{s}$ the resulting series
are numerically consistent with series in even powers of $\tau$ only.
We have also noted a similar simplifying role played by $\sqrt{s}$
in the ``short-distance'' terms which ultimately trace back to the Toeplitz
determinant giving the diagonal correlations $C(N,N)$ and to the
quadratic recursion relations for general $C(M,N)$.\footnote{See also
the discussion of the quadratic recursion relations in Itzykson and
Drouffe \cite{id} where rescaling by $\sqrt{s}$ was used to simplify the
scaling limit.} This result is not entirely unexpected. The
singular part of the free energy in zero field is an even function of
$\tau$ and the magnetization ${\mathcal M} = (1-s^{-4})^{1/8}$ is $s^{-1/4}
(-\tau)^{1/8}$ times an even function of $\tau.$ Nonlinear scaling field
analysis then predicts, in the absence of corrections, that the
``scaling'' part of the susceptibility is $s^{-1/2}|\tau|^{7/4}$ times
an even function of $\tau$. Thus although our results for $F_{\pm}$
are not
consistent with the complete absence of correction terms as discussed
in section 7 the prediction that $\sqrt{s}F_{\pm}$ is even in 
$\tau$ does appear to be preserved to all orders.

To determine the coefficients in $F_{\pm}$ we return to the
unscaled $\chi_{\pm}$ of eqs.~(\ref{1a},\ref{1b}) so that the terms in
$F_{\pm}$ are now singularities of the function and contribute to the high
order coefficients in the series. The cases $T<T_c$ and $T>T_c$ are again
similar; for $T<T_c$ the unwanted contributions from the
``short-distance'' part of $\chi$ are 
subtracted as in the $s$-plane analysis described in
section \ref{sect:anal}.
 We use essentially the same
smoothing except for a $1/4$ shift in power necessitated by the
difference in the singularity structure generated by the 
$\chi$ rescaling of section \ref{sect:anal}.
That is, we replace the remainder eq.~(\ref{717}) by

\begin{equation}\label{719}
R^-_n = n^{1/2} D_b^4 n^{-3/4} (n^4 D_a^4)^3 n^{13/4} K^-_n
\end{equation}
and reduce the remainders $R^-_n$ in (\ref{719}) by a least squares fitting
to the unknown $F_{-}$ coefficients in a procedure similar to that
described in \cite{nicb}. Fitting intervals $\Delta n > 128$ are typically
used, and an FFT of the residuals is very useful as
a diagnostic to interpret the observed oscillations in the residuals
in terms of $\chi$ singularities on the circle $|s^{-1}|=1$. 
Because the highest order terms in $F_{-}$ are
not fixed (i.e. known) unlike the ``short-distance'' terms, they tend to
float and become effective amplitudes that incorporate all the higher
order effects including the ``short-distance'' contributions that have not been
subtracted. One technical result of this is that the residuals we
observe in (\ref{719}) are some six orders of magnitude smaller than the
residuals we obtained in (\ref{717}). In part this means that whereas
(\ref{717}) was more than adequate as a smoothing operation, (\ref{719}) is
marginally so. In addition, we observe here for the first time one of
the singularities from $\ch{6},$ 
and using the cut information given in eqs.~(\ref{new27}, \ref{new28}) can
subtract it as

\begin{eqnarray}\label{720}
 K^-_{2m} &\rightarrow & K^-_{2m} + 7^8 15/(16\pi \hat{m}^{35/2})
(3 \sqrt{7}/\pi^2)^{3/4}\{\cos(2\hat{m}\arccos(3/4) -\pi/8) \\ \nonumber
& - & 215\sqrt{7}/(32\hat{m})\sin(2\hat{m}\arccos(3/4) -\pi/8)
 + {\rm O}(1/\hat{m}^2)\},
\end{eqnarray}
 where $\hat{m} = m-1/4$.

A similar analysis has been carried out for $T>T_c$ and our
estimates for the coefficients of $F_{\pm}$
are given in the Appendix.
The coefficients through O$(\tau^5)$ are unambiguously rational and have
been fixed in the final fittings. We have also set to zero all
coefficients of odd powers of $\tau$ in the product $\sqrt{s}F_{\pm}$
to O$(\tau^{15})$ but have allowed variable coefficients of $\tau^{16},$ 
$\tau^{17}$ and
$\tau^{18}.$ These latter coefficients are, as expected, sensitive to
whether we stop the ``short-distance" subtraction at O$(\tau^{14})$ or add in
the additional terms we have estimated by Pad\'e methods.
The terms quoted in
the Appendix on the other hand are completely stable and thus we
believe reliable except possibly for the last digit.

When we relax the constraint of zero amplitude on individual odd
$\tau^k$ terms in $\sqrt{s}F_{\pm}$ for integer $5\le k \le 15$ we
find no significant improvement in our fits and the resulting 
amplitudes are consistent with zero. For example, when $T<T_c$, we 
find best fit coefficients of $\tau^k$, $k$ odd, that in absolute 
magnitude are all less than $\approx 4 \times 10^{2k-33}$. For 
$T>T_c$ the corresponding bounds can be as much as $100$ times larger.
But in both cases these bounds are of a magnitude similar to what we 
estimate is the uncertainty in the even order coefficients given in 
the Appendix. Thus we believe $\sqrt{s}F_{\pm}$ even in $\tau$ to be an exact 
symmetry of the scaling-amplitude function.

An alternative analysis using the traditional variable $v = \tanh{K}$ was
also carried out. The natural boundary singularities at $|s|=1$ are
mapped to two circles $|v \pm 1| = \sqrt{2}$. 
In this expansion variable, the ferromagnetic and anti-ferromagnetic
critical points are at $v = \pm(\sqrt{2} - 1)$ respectively, and
all other points on the two circles are farther
away from the origin. Hence the amplitudes of
any other singularities are exponentially damped and may be neglected in the
analysis. The two analyses are in complete agreement, but the more
detailed $s$-plane analysis provides greater precision. For contemplated
future analyses on other lattices, for which less is known about the
natural boundary singularities, it may be necessary to
use the $v$-plane analysis.

\section{Expected scaling form of the susceptibility.}\label{sect:scaling}

The basic scaling Ansatz for the singular part of the free-energy of
the two-dimensional Ising model is
\begin{eqnarray}\label{s1}
f_{\rm sing}(g_t,g_h,\{g_{u_j}\}) &=& -g_t^2\log{|g_t|}\tilde{Y}_{\pm}
(g_h/|g_t|^{y_h/y_t},\{g_{u_j}|g_t|^{-y_j/y_t}\}) \nonumber \\
& + & g_t^2Y_{\pm}(g_h/|g_t|^{y_h/y_t},\{g_{u_j}|g_t|^{-y_j/y_t}\}).
\end{eqnarray}
Here $g_t, g_h, g_{u_j}$ are nonlinear scaling fields associated with the
thermal field $t,$ the magnetic field $h$ and the irrelevant fields $\{u_j\}.$
The exponents $y_t, y_h > 0$ are the thermal and magnetic exponents, and
$y_j < 0$ are the irrelevant exponents.\footnote{For the two-dimensional Ising
model, $y_t = 1$ and $y_h = \frac{15}{8}$. This scaling Ansatz assumes only a
single resonance, between the identity and the energy. That the dimensions are
integers implies higher powers of $\log{t}$. The surprise is the power of $t$
at which higher powers of $\log{t}$ enter. We thank Andrea Pelissetto for
this clarification.}. The nonlinear scaling fields have expansions
\begin{eqnarray}\label{s2}
g_t & = & \sum_{n \ge 0} a_{2n}(t,u)h^{2n}, \mbox{ } a_0(0,u)=0,
 \nonumber \\
g_h& = & \sum_{n \ge 0} b_{2n+1}(t,u)h^{2n+1}, \nonumber \\
g_{u_j}& = & \sum_{n \ge 0} c_{2n}(t,u)h^{2n},
\end{eqnarray}
where $a_{2n}, b_{2n+1}, c_{2n}$ are smooth functions of $t$ and
$u \equiv \{u_j\}.$

If irrelevant fields are neglected, then the known zero field
free energy forces the equalities $\tilde{Y}_+(0) = \tilde{Y}_-(0)$
and $Y_+(0) = Y_-(0)$. Furthermore, the absence of logarithmic terms in
the known magnetization and the divergent part of the susceptibility
requires the derivatives $\tilde{Y}'_ {\pm}(0)$ and $\tilde{Y}''_{\pm}(0)$
to vanish. Aharony and Fisher have conjectured~\cite{AFb}, almost certainly
correctly, that there are no logarithms multiplying the leading power
law divergence of all higher order field derivatives, in which case
the $\tilde{Y}_{\pm}$ are constants and analyticity on the critical
isotherm for $h \ne 0$ demands $\tilde{Y}_+ = \tilde{Y}_-.$
With all these constraints built in, one can show the scaling 
Ansatz~(\ref{s1}) together with the field expansions~(\ref{s2}) lead to
\begin{eqnarray}\label{s3}
f(t,h=0) & = & -A(a_0(t))^2\log{|a_0(t)|} + A_0(t), \nonumber \\
{\mathcal M}(t<0,h=0) & = & B b_1(t)|a_0(t)|^{\beta}, \\
\beta^{-1}\chi_{\pm}(t,h=0) & = & C_{\pm}(b_1(t))^2|a_0(t)|^{-\gamma} - 
E a_2(t)a_0(t)\log{|a_0(t)|} + D(t),
\end{eqnarray}
where $A,$ $ B,$ $C_{\pm}$ and $E$ are constants and 
$\beta=1/8,$ $\gamma=7/4.$ The
free energy and magnetization eqs.~(\ref{s3}) determine the scaling field
coefficients $a_0(t)$ and $b_1(t)$ in~(\ref{s2}). The presence of irrelevant
scaling fields will be expected to manifest themselves as deviations
from the predicted form of the susceptibility in~(7.4) and/or as
deviations from the unique prediction for the coefficient of $C_{\pm}.$

Working as usual in the temperature
variable $\tau = (1/s - s)/2$, we write the predicted
isotropic susceptibility from~(7.4) as
\begin{equation}\label{7.5}
\beta^{-1}\chi_{\pm}(\tau,h=0) = C_{0_\pm}(2K_c\sqrt{2})^{7/4}|\tau|^{-7/4}
F({\rm A\&F})
- {E}_{0}/(2K_c\sqrt{2})\tau\log{|\tau|}e_{0}(\tau) + D_0(\tau)
\end{equation}
where $F({\rm A\&F})$ has already been given in eq.~(\ref{9c}). The
``short-distance" contribution to $\chi$ is here predicted to be given by the
sum of the term containing $e_0(\tau),$ which arises as the mixing of
the first two terms in the expansion of $g_t$ in~(\ref{s2}), and the
analytic $D_0(\tau).$

The clear implication of eq.~(\ref{9d}), which shows the exact
$F_{\pm}$ is not equal to $F({\rm A\&F})$ is that irrelevant variables
do play a role, and the multipliers $ F_{\pm}$
represent the contribution of a number of scaling fields.
While there are suggestions in the literature for what
these scaling fields might be\footnote{
 An excellent early discussion of the types of corrections that might
 be found, together with a search for some of them can be found in
 Bl\"{o}te and den Nijs \cite{BdenN}. Developments in our
understanding of the predictions of conformal field
theory~\cite{cardya,cardyb,cardyc}
lead us to believe that a virtually complete explanation of
corrections to scaling is obtainable, at least in principle.
Our analysis supports the conclusion of Barma and Fisher\cite{FB}
that a correction-to-scaling term with exponent $\theta = 4/3$ is
absent for the pure $S=1/2$ Ising model susceptibility considered here.
 A mechanism in terms of generators of the energy family
of the Virasoro algebra is adduced by Caselle et al. \cite{Pe00}
which gives rise to corrections at order $\tau^4$ as observed.}
it is
unlikely that a unique identification could be made here since we are
dealing with a single isolated model with no free parameters to vary.
A corresponding analysis of the anisotropic square, and the
 triangular and hexagonal lattices is
likely to be enlightening in this regard. In \cite{ap87} a study
of difference equations is given, which implicitly outlines what
is needed to obtain the difference equations for the hexagonal and
triangular lattices. Additional material in this respect can also
be found in \cite{AJP00}.

Note however that two different effects manifest themselves.
At fourth order in $t$ (or, equivalently, $\tau$)
scaling under the assumption of
only two nonlinear scaling fields breaks down, as evidenced by
the difference between the coefficients of $\tau^4$ in
eqs.~(\ref{9d}) and~(\ref{9c}).
However the corresponding high- and low-temperature amplitudes still satisfy
$C_0^+/C_0^-=C_j^+/C_j^-$ for all $j \le 5.$ For $j > 5,$ not only
does simple scaling fail to hold, but this equality also breaks down.

In the vicinity of the anti-ferromagnetic point in the high-temperature
phase, $\chi$ is given exclusively by a ``short-distance'' term (\ref{6e}).
Both $B_{\rm af}$ and $B_{\rm f}$ from eqs.~(\ref{6e}) and~(\ref{9b})
have expansions of the same form~(\ref{6f}),
where the sum over $p$ is restricted to $p^2\le q.$
The coefficients
in this expansion can be determined from the short-distance
correlations, and accurate values for the expansions of $B_{\rm f/af}$ are
given in the Appendix through {\rm O}$(\tau^{14}).$

Again we note that there are terms in these
``short-distance'' functions that are not of the Aharony and Fisher
\cite{AFa,AFb} predicted form~(\ref{7.5}) based on the absence of irrelevant
variables.

We conclude with some speculative remarks.
We cannot account physically
for terms of order $t^q(\log|t|)^p,$ with $p \ge 2$ and $q \ge p^2$ in $B_f$,
though we can see their origin mathematically, as discussed below
eq.~(\ref{7b}) and in footnote 12.
While terms without logarithms and terms of order $t^q(\log|t|)$ are expected,
it is surprising (to us) that higher powers of $\log|t|$ enter at the orders
they do. 

From conformal field theory, we have predictions for the irrelevant exponents
$y_j = -2, -4, -6, \ldots.$
The fact that $f_{+}^{(k)}=f_{-}^{(k)}$
for $k < 6,$ though $f_{\pm}^{(k)}$ is not equal to the corresponding term
in~(\ref{9c}) for $k=4, 5$ suggests the presence of only a {\bf single}
irrelevant operator contributing at order $\tau^4,$ while the breakdown of
high-low temperature symmetry in $F_{\pm}$ at O$(\tau^6)$ suggests that more
than one scaling operator couples to the lattice magnetization at this order.
A corresponding study to that reported here on the triangular and
honeycomb lattices, as well as on the anisotropic square lattice is
likely to be enlightening, and we hope to report on this in future.

{\bf Acknowledgments}
We would like to thank Larry Glasser, Mireille Bousquet-M\'elou and
John Cardy for many helpful discussions
and calculations, Andrea Pelissetto for adding to our understanding of scaling
theory and the provision of a preprint, 
Barry McCoy, Alan Sokal and Michael Fisher for invaluable comments
on earlier versions of the manuscript, and Douglas Abraham for useful
comments on the history of the Ising model.
AJG and WO would like to thank the Australian Research Council for financial
support, and JHHP thanks the NSF for support in part by
NSF Grant no. PHY 97-22159.
\section*{Appendix}
Here we list the expansion coefficients $b_{p,q}$
of the ``short-distance" functions $B_{\rm f/af}$
defined in~(\ref{6f}) to O$(\tau^{14})$.
The prefactor $(\sqrt{1 + \tau^2}+\tau)^{1/2} = 1/\sqrt{s}$
is understood to be expanded in a series in 
$\tau.$ The leading constant and coefficient of
$\tau \log{|\tau|}$ have been reported previously~\cite{Kong}.
The $F_{\pm}$ series as deduced in section~\ref{sect:ca} are also given.

\begin{align*}
&B_{\rm f} = {(\sqrt{1 + \tau^2}+\tau)}^{1/2}\\
&[- 0.104133245093831026452160126860473433716236727314 \\
   & - 0.07436886975320708001995859169799500328047632028\tau \\
   & - 0.0081447139091195995371542858655723893266057740\tau^2 \\
   & + 0.004504107712232015926355020852986970591364528\tau^3 \\
   & + 0.23961879425472180967837072450742931180586109\tau^4 \\
   & - 0.0025399505953392329612162686121616238176205\tau^5 \\
   & - 0.235288909669962491804066210882350821445764\tau^6 \\
   & + 0.00191570753170091141409998516033460855797\tau^{7} \\
   & + 0.2143400966115384518711435705343125612378\tau^{8}\\ 
   & - 0.000883215706003328768611915486246075323\tau^{9} \\
   & - 0.19422062840719623752953468278129284679\tau^{10} \\
   & + 0.0000072335097772632765778839359680528\tau^{11} \\
   & + 0.177102037555467190714704023746648559\tau^{12} \\
   & + 0.0006888110962684387331860926084517\tau^{13} \\
   & - 0.16279253648974618861881216566686\tau^{14}\\
   & + \log|\tau|\\
&  ( 0.032352268477309406090656526721221666637730948898\tau \\
   & - 0.0057755293796884630091487564013201013677152980\tau^3 \\
   & + 0.059074961290345476578516085774495545264759330\tau^4 \\
   & + 0.00305849157585622544005057759535229287938174\tau^5 \\
   & - 0.0591662722088409053375931018028970139567911\tau^6 \\
\displaybreak[4] \\
   & - 0.002067088393167114141650194740281136875636\tau^{7} \\
   & + 0.05424693070421409615112542698595864778919\tau^{8} \\
   & + 0.0010601025315498815900774416057541651837\tau^{9} \\
   & - 0.049300253157082567741316861339709144063\tau^{10} \\
   & - 0.00026830064161204203467706137637400358\tau^{11} \\
   & + 0.0450270525719569186212816103126356308\tau^{12} \\
   & - 0.00034332683257234543036792535081332\tau^{13}\\
   & - 0.041428586463052869356803144137620\tau^{14})\\
& + (\log|\tau|)^2 \\
&  ( 0.0093915698711458721317953318727075770649513654\tau^4 \\
   & - 0.00869592546287923802156416645191752987912922\tau^6 \\
   & + 0.007669481493104540876445085447422616885330\tau^{8} \\
   & + 0.00015428438297902275440225213783285077606\tau^{9} \\
   & - 0.0068054076881441249098452112921129773269\tau^{10} \\
   & - 0.000310520937481414524040686012525223279\tau^{11} \\ 
   & + 0.00611386643219454473116391019937140965\tau^{12} \\
   & + 0.000444606198235804033861443998682830\tau^{13}\\
   & - 0.0055571002151161308034896964314679\tau^{14})\\
&+(\log|\tau|)^3 \\
 & ( - 0.000015771569138451840480001012621461738178\tau^{9} \\
   & + 0.0000344282066208887553647799856857753380\tau^{11}\\
   & - 0.0000524427177487226174161583779149393\tau^{13})],
\end{align*}
\begin{align*}
&B_{\rm af}  = {(\sqrt{1 + \tau^2}+\tau)}^{1/2}\\
&  [ 0.1588665229609474882333592313690210116925239008416 \\
   & + 0.149566836938535905194382029433591286374711207262\tau \\
   & + 0.01071222587983288033470968550659996768542030678\tau^2 \\
   & + 0.0127530188399624019539552078052153609134674971\tau^3 \\
\displaybreak[4] \\
   & - 0.011741188869656263932121387296300743594029390\tau^4 \\
   & - 0.01406604087566590060620992322775625815515533\tau^5 \\
   & + 0.0131064546156258402249424759665220798848681\tau^6 \\
   & + 0.012239696625538370626786005459530159711716\tau^{7} \\
   & - 0.01184019404541084813958321002995560636877\tau^{8} \\
   & - 0.0105854093023116661362507232392645147231\tau^{9} \\
   & + 0.010151560037724359473553197335905170854\tau^{10} \\
   & + 0.00908000411233119371610549453140718281\tau^{11} \\ 
   & - 0.0085420122287896456879459087815054030\tau^{12} \\
   & - 0.00771702694013238358077176900242074\tau^{13} \\
   & + 0.007123677682511208149032476379667\tau^{14}\\
& + \log|\tau| \\
 & ( - 0.1553171901580110585934133538932734529992121600305\tau \\
   & + 0.03206714814586975221843437287457551882247161782\tau^3 \\
   & - 0.0077168875724615093064542922101962689299768599\tau^4 \\
   & - 0.015675211573817078943430169665269657287132910\tau^5 \\
   & - 0.00028554245153720354627897919087710530890677\tau^6 \\
   & + 0.0096072545027321808179041903535130201897217\tau^{7} \\
   & + 0.004835406420625092236673413378307375358908\tau^{8} \\
   & - 0.00606499034448050751379194815071149626812\tau^{9} \\ 
   & - 0.0073400150414474023562454611060746360875\tau^{10} \\
   & + 0.003910356521403913091050321141297009252\tau^{11} \\
   & + 0.00870842744568158003036434762719697635\tau^{12} \\
   & - 0.002697783010884752101384006121375890\tau^{13} \\
   & - 0.0094056230380765607719474925088649\tau^{14})\\
& +(\log|\tau|)^2 \\
  &( 0.01153371437882328027949011442761203640684043805\tau^4 \\
   & - 0.011311734920691560067535056532207842716405684\tau^6 \\
\displaybreak[4] \\
   & + 0.0100457687111988577404299867962466051265974\tau^{8} \\
   & - 0.000475698571097159420906450182271928179428\tau^{9} \\
   & - 0.00878397202228689639470985683437717938463\tau^{10} \\
   & + 0.0011571801729636538264100914359800686355\tau^{11} \\
   & + 0.007680651109512704070606639646988801296\tau^{12} \\
   &- 0.0018650912616201532939412831153215046\tau^{13} \\
   & - 0.00674470189451526288478200059343432\tau^{14}) \\
& + (\log|\tau|)^3 \\
 & ( 0.0000578997194764877297760067221144062249541\tau^{9} \\
   & - 0.00016991508824012890240796446744935908812\tau^{11}\\
   & + 0.00032664884687465587957270016883093909\tau^{13})]
\end{align*}
\begin{align*}
 F_+ &= 1 +\tau/2 +5\tau^2/8 +3\tau^3/16 -23\tau^4/384 -35\tau^5/768
         -0.1329693327418753330\tau^6 \\
& -0.05899768720427100\tau^7 
         +0.121586869804903\tau^8 +0.0766007994119\tau^9 \\
&     -0.10751871874\tau^{10} -0.078346589\tau^{11}
 +0.0960583\tau^{12}
         +0.07592\tau^{13}\\
& -0.087\tau^{14} -0.1\tau^{15}+ \ldots\\
& = (\sqrt{1+\tau^2}+\tau)^{1/2} (1 +\tau^2/2 -\tau^4/12
          -0.1235292285752086663 \tau^6\\
& +0.136610949809095 \tau^8 -0.13043897213 \tau^{10}
 +0.1215129 \tau^{12} -0.113 \tau^{14} + \ldots )
\end{align*}
\begin{eqnarray*}
 F_- &=& 1 +\tau/2 +5 \tau^2/8 +3 \tau^3/16 -23 \tau^4/384 -35 \tau^5/768
         -6.330746944662603289734 \tau^6  \\
	 &-&3.1578864931646349782 \tau^7
         +5.46225118896595954 \tau^8 +3.521655160482472 \tau^9  \\
        & -& 4.6602157191837 \tau^{10} -3.40963923001 \tau^{11}
     +4.055875878 \tau^{12} +3.2008085 \tau^{13} \\
&- &3.59746 \tau^{14} -2.985 \tau^{15} +\ldots\\
& =&{(\sqrt{1 + \tau^2}+\tau)}^{1/2} (1 +\tau^2/2 -\tau^4/12
          -6.321306840495936623067 \tau^6 \\
&+&6.25199747046024329 \tau^8  
          - 5.6896599756180 \tau^{10} +5.142218271 \tau^{12}\\
&-& 4.67472 \tau^{14} + \ldots)
\end{eqnarray*}
The last digit in each term above may not be reliable.

\noindent
For completeness we give also the leading susceptibility
amplitudes evaluated to higher accuracy than reported in \cite{nica}:
\begin{eqnarray*}
C_0^+ &\!=\!& 1.000815260440212647119476363047210236937534925597789(2K_c
 \sqrt{2})^{-7/4} \sqrt{2} \\
C_0^- &\!=\!& 1.000960328725262189480934955172097320572505951770117(2K_c
 \sqrt{2})^{-7/4} \sqrt{2}/(12 \pi).
\end{eqnarray*}

\end{document}